\documentstyle[11pt,aaspp4,flushrt,psfig]{article}

\def\ie{{i.e.}}

\def\etal{et al.}
\def\simleq{\stackrel{<}{\scriptstyle\sim}}
\def\simgeq{\stackrel{>}{\scriptstyle\sim}}
\def\*{{$^*$}}

\newcommand{\uv}{\mbox{$u$-$v$}}

\newcommand{\sn}[1]{\mbox{s$^{#1}$}}
\newcommand{\Msol}{\mbox{M\raisebox{-.6ex}{$\odot$}}}
\newcommand{\kms}{\mbox{km s$^{-1}$}}
\newcommand{\muas}{\mbox{$\mu$as}}
\newcommand{\muasyr}{\mbox{$\mu$as~yr$^{-1}$}}
\newcommand{\masyr}{\mbox{mas~yr$^{-1}$}}

\lefthead{Bietenholz \etal}
\righthead{One-sided jet in M81}

\begin{document}
      
\title{A Stationary Core with a One-sided Jet in the Center of M81}

\author{M. F. Bietenholz and N. Bartel}
\affil{Department of Physics and Astronomy, York University, Toronto, M3J~1P3, Ontario, Canada}
 
\author{and M. P. Rupen}
\affil{National Radio Astronomy Observatory, Socorro, New Mexico 87801, USA}

\vspace{0.2in}
\centerline{(accepted for publication in the {\em Astrophysical Journal})}

\begin{abstract}

The nucleus of the nearby spiral galaxy M81 was observed at 8.3~GHz
with a global VLBI array at 20 epochs over four and a half years, with
a linear resolution at the source of about 2000~AU or
0.01~pc. Phase-referenced mapping with respect to the geometric center
of supernova 1993J in the same galaxy enabled us to find, with a
standard error of about 600~AU, a stationary point in the source
south-east of the brightness peak. We identify this point as the
location of the core and the putative black hole at the gravitational
center of the galaxy.  The $2\sigma$ upper bound on the core's average
velocity on the sky is $\leq 40~\muasyr$ or $\leq 730$~\kms\ (relative
to the center of SN1993J, in excess of galactic rotation).  A short,
one-sided jet extends towards the north-east from the core, in
projection approximately in line with the rotation axis of the galaxy,
and towards the well-known extended emission 1~kpc further out.  The
orientation of the jet varies smoothly, with timescales of about one
year, and an rms of 6\arcdeg, around its mean position angle of
50\arcdeg.  Occasionally the jet appears to bend to the east.  The
length of the jet is only about 1~mas (3,600~AU), and varies with an
rms of about 20\% from epoch to epoch. The inferred speeds are below
$0.08 c$.  The total flux density of the core-jet varies erratically,
changing on occasion by a factor of two over a few weeks, without any
significant changes in the source size and orientation.  The inferred
velocity of the plasma flow is $>0.25 c$.  The results are consistent
with a model in which plasma condensations with short lifetimes are
ejected relativistically from the core on a timescale of less than a
few weeks. The condensations travel along a tube whose pattern and
geometry are also variable but only on a timescale of about one
year. The central engine of M81 has qualitative similarities to those
of powerful active galactic nuclei of radio galaxies and quasars, and
may also represent in power and size a scaled-up version of the
largely hidden nucleus in our own Galaxy.

\end{abstract}

\keywords{galaxies: individual (M81) --- galaxies: nuclei --- radio continuum: galaxies}

\section{Introduction}

The nearby galaxy M81 (NGC 3031, 0951+693) is a grand-design spiral
that resembles our own Galaxy in type, size, and mass. Like our
Galaxy, it contains a nuclear radio source that is most likely
associated with a supermassive black hole in the gravitational center
of the galaxy.  M81 also shares some characteristics with radio
galaxies and quasars.  Its nucleus has a high X-ray luminosity ($\sim
1.7 \times 10^{40}$~erg~s$^{-1}$, Elvis \& Van Speybroeck 1982) and a
non-stellar UV continuum (Ho, Fillipenko \& Sargent 1996).  Optical
spectroscopy shows broad H$\alpha$ lines (Peimbert \& Torres-Peimbert
1981) and double peaked broad other emission lines, which in galaxies
with an active galactic nucleus (AGN) are almost always associated
with jets (Bower \etal\ 1996).  The central radio source in our
Galaxy, Sgr~A\*, is largely hidden behind scattering clouds of gas
that have so far allowed only the crudest determinations of its
intrinsic size and orientation (Lo \etal\ 1998; Krichbaum \etal\
1998).  M81's central radio source, on the other hand, is virtually
unaffected by scatter broadening at most radio frequencies.  M81 is
also, at a distance of 3.63~Mpc (Freedman \etal\ 1994), the nearest
spiral galaxy with a central compact radio source, and, with the radio
galaxy Cen~A (Tingay \etal\ 1998), the nearest galaxy with an AGN
altogether.  With a radio luminosity of $10^{37.5}$~ergs~\sn{-1}, the
nucleus of M81 could therefore be both a more readily observable
analog to Sgr~A\*, and a useful link between our galactic center and
the more powerful nuclei of radio galaxies and quasars.

Previous observations have confirmed that M81\*\footnote{Following
Melia (1992) and others we adopt the convention of adding an asterisk
to the name of the galaxy when referring to the VLBI source in the
galaxy's center.} is similar to typical AGN in terms of its
very-long-baseline interferometry (VLBI) brightness temperature of
$\sim10^{10}$~K at 8~GHz and its radio spectrum, which is slightly
inverted.  Its spectral index, $\alpha$ ($S \propto \nu^\alpha$), is
about +0.2 from a frequency of $\sim 1$~GHz to its turnover frequency
of $\sim 100$~GHz (Reuter \& Lesch 1996; Bartel \etal\ 1982, B82
hereafter).  The high turnover frequency suggests an unusually small
source; this is consistent with VLBI observations which gave a size of
700~AU $\times$ 300~AU at 22~GHz (Bietenholz \etal\ 1996, hereafter
Paper~I; see also Bartel, Bietenholz \& Rupen 1995, Bietenholz \etal\ 
1994, B82, and Kellermann \etal\ 1976).  It is larger at lower
frequencies, with the length of the major axis being proportional to
$\nu^{-0.8}$ between 2.3~GHz and 22~GHz.  The orientation of M81\* is
also frequency dependent, bending from $\sim$40\arcdeg\ at 22~GHz to
$\sim$75\arcdeg\ at 2.3~GHz. This result, combined with a slight
asymmetry in the brightness within the source with the source being
fainter towards the north-east (NE), suggests a typical, albeit tiny,
core-jet source with a bent jet.
One surprise is that no significant changes in the size and
orientation of M81\* were seen between the epochs of the observations
in 1981 and 1993 (Paper~I, B82), giving an extremely low nominal
expansion velocity of $-60 \pm 60$~\kms.  Such low velocities are
unusual for core-jet sources where plasma condensations are
ejected relativistically. They can perhaps be understood if the plasma
condensations are ejected relativistically from the core, but then
follow nearly identical trajectories, and are ejected frequently
enough to keep the integrated flux distribution fairly constant.  Here
we report on observations from 20 epochs of VLBI observations of
M81\*, taken between 1993 and 1997.  Using phase referencing, these
observations have allowed us to identify astrometrically a small
stationary region in the south-west (SW) of the source as the probable
core, and to detect and characterize clear structural variability in
the jet.

We describe the observations and data reduction in sections
\S\ref{obss} and \ref{datreds}.  In \S\ref{fluxds} we give the radio
lightcurve determined with the NRAO\footnote{The National Radio
Astronomy Observatory is a facility of the National Science Foundation
operated under cooperative agreement by Associated Universities, Inc.}
Very Large Array (VLA).  In \S\ref{vlbmaps} we present six
representative images illustrating the structural variability.  In
\S\ref{vlbmods} we examine this variability in more detail by fitting
one- and two-component models in the \uv~plane.  In \S\ref{core} we
give our astrometric results from phase-referencing to supernova
1993J, and identify the stationary core in the SW and the variable
jet.  In \S\ref{jet} we discuss the properties of the jet.  In
\S\ref{discuss} we discuss our results in terms of a core-jet model
for M81\*, and finally, in \S\ref{concss} we summarize our
conclusions.  (For a preliminary presentation of some of these
results, see Bietenholz, Bartel \& Rupen 1997).

\section {Observations \label{obss}}

The observations were made with a global array of between 11 and 18
telescopes (see Table~\ref{antab}) with a total time of 12 to 18 hours
for each run, giving us exceptionally good \uv~coverage.  Each
telescope was equipped with a hydrogen maser as a time and frequency
standard. The data were recorded with the VLBA (Very Long Baseline
Array) and either the MKIII or the MKIV VLBI systems with sampling
rates of 128 or 256~Mbits per second.  M81\* was observed as a phase
reference source for the continuing multi-frequency program of VLBI
observations of SN1993J, located $\sim170\arcsec$ away towards the
south south-west in a spiral arm of the galaxy (Bartel \etal\ 1999,
1994).  In general, a cycle time of $\sim3$~min was used, in which
M81\* was observed for 70s and SN1993J for 120~s (a longer cycle time
was used only during the first three epochs).  In addition the sources
OQ208 and 0954+658 were observed occasionally during each 
session as fringe finders and calibrator sources.  In each session
data at two to four frequencies out of a total of six (1.7, 2.3, 5.0,
8.3, 15, and 22~GHz) were recorded, with 5.0 and 8.3~GHz being the
standard frequencies used in (almost) every session. Here we report on
the VLBI observations at 8.3~GHz only, where we always recorded right
(IEEE convention), and for later runs also left circular polarization.

The phased VLA was used in all of these VLBI observations, and the
(simultaneous) interferometric data from it provided accurate flux
measurements at 8.3~GHz of M81 and SN1993J.  Both senses of circular
polarization were recorded with a bandwidth of 50~MHz per
polarization.  During the four and a half years of observations data
were taken in all four standard, as well as in various hybrid VLA
configurations. The dates of observations and the array configurations
are given in Table~\ref{fluxt}.

\section {Data Reduction \label{datreds}} 

The VLBI data were correlated using the VLBA processor in Socorro, NM.  Further data
reduction, \ie\ fringe-fitting, editing, and initial calibration, was
done with NRAO's software package, AIPS, in the usual way.  The final
calibration was done by starting with a single Gaussian model and then
iteratively self-calibrating the complex antenna gains. Phases were
derived first, with a solution interval of $\sim 1$~min, and then
amplitudes, with final solution intervals of 1 to 2~h.

In order to more precisely describe the structural properties of the
nucleus, we turned to model-fitting.  We first averaged the data for
each epoch into scans of $\sim 1$~min duration. We then used the AIPS
least-squares fitting program OMFIT which simultaneously fits the
time-varying antenna gains (self-calibrates) and solves for the model
parameters.  Solving again for the antenna gains at the same time as
solving for the source parameters is essential because the two will
generally be correlated.  The resulting model parameters were derived
in conjunction with antenna gains which had solution intervals of
1~min in phase and 1~h in amplitude.
 
Total flux densities were obtained from the interferometric VLA data
from the M81\* pointings of our observing runs.  In particular, we
derived the flux densities from CLEAN images made from the fully
phase-selfcalibrated and naturally weighted data. The flux densities
are on the scale of Baars \etal\ (1991).  In all cases the internal
uncertainties in the VLA flux densities, including the possible effect
of extended structure\footnote{There is a small amount of extended
emission near M81\* on arcsecond scales which in principle could
contaminate our results from compact array configurations. However,
the total flux density of any such emission is less than a few percent
of that of M81\* (Bartel \etal\ 1995; Kaufman \etal\ 1996), excluding
it as a source of significant error.}, were dominated by the standard
error in the VLA calibration, which we conservatively assume to be
5\%.

\section {VLA total flux densities\label{fluxds}}

\begin{figure}
\plotfiddle{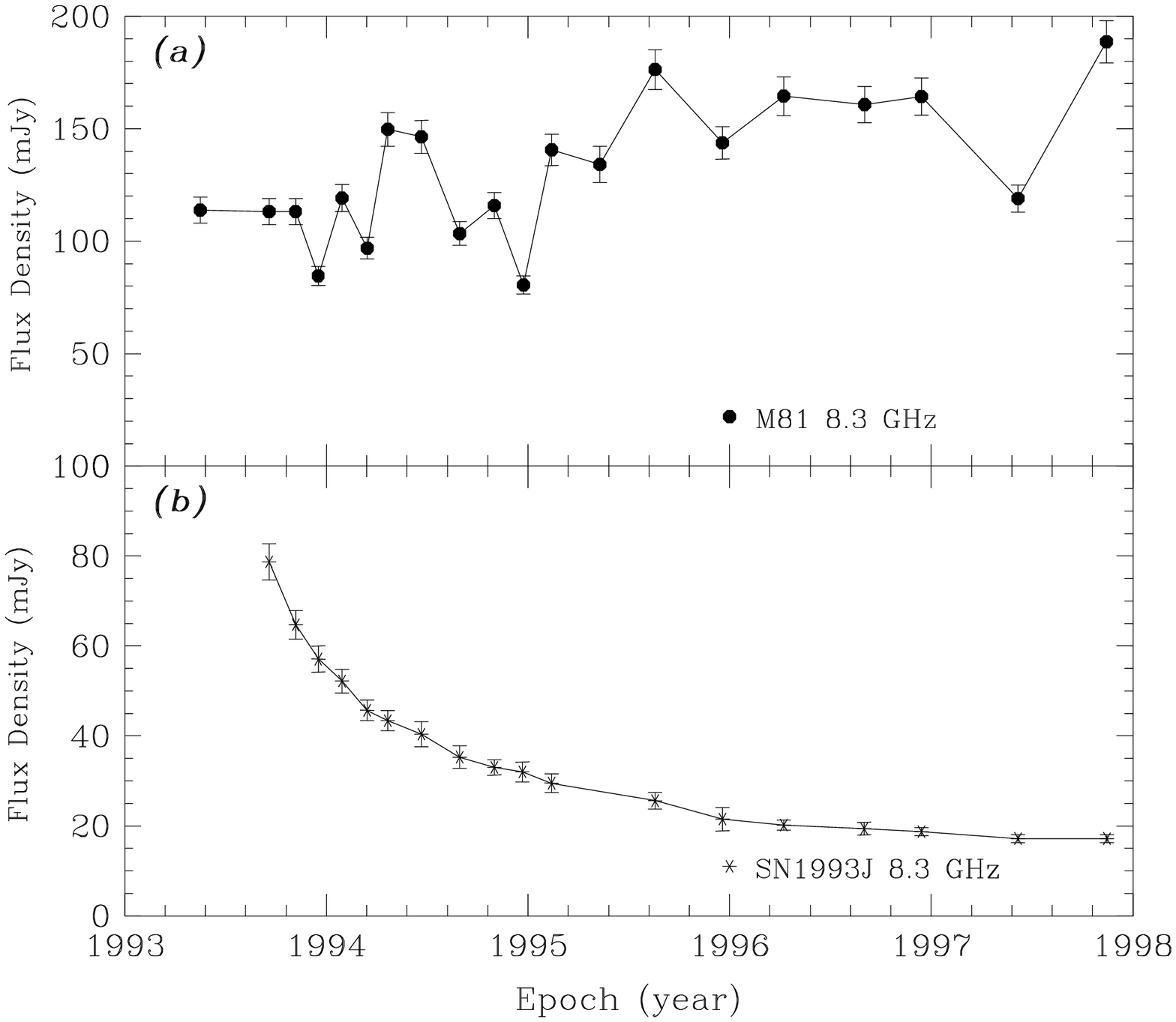}{5in}{0}{70}{70}{-215}{-180}
\figcaption{Total flux densities at 8.3~GHz derived from VLA imaging
observations.  In $(a)$ we show the total flux density evolution
of M81\*.  For comparison, in $(b)$ we show the total flux density
evolution of SN1993J which was obtained from the same imaging runs and
which shows only the expected smooth decline, and does not exhibit the
variability shown by M81\*.\label{fluxp}}
\end{figure}

We list the total flux densities of M81\* at 8.3~GHz and their mean
and rms (root-mean-square) scatter about the mean for 20 epochs in
Table~\ref{fluxt}. These same flux densities are graphed in
Figure~\ref{fluxp}$a$.  For comparison, we show in
Figure~\ref{fluxp}$b$ the flux densities of SN1993J, which were
derived from SN1993J pointings which alternated with M81\* pointings
in the same observing runs, and were calibrated using the same gain
amplitudes, and so any calibration problems in the M81\* data would
almost certainly be apparent in the flux densities measured for
SN1993J. Since, however, the latter show only the expected smooth
decline to well within the errors, we are confident that our
measurements of the variable flux densities of M81\* are reliable and
correct within the quoted errors.  All our uncertainties are one
standard error unless otherwise noted.

The flux densities of M81\* fluctuate strongly.  They vary by over a
factor of 2 from 81~mJy to 189~mJy, with the rms deviation being 23\%
of the mean of 127~mJy. The variation appears random from epoch to
epoch.  The degree of variability and the mean flux density are
consistent with those reported by others (Ho \etal\ 1999; Crane,
Guiffrida \& Carlson 1976).

We examined our data for possible flux density variations during the
course of an observing run.  No significant variations were found ---
For the 12 epochs we inspected, we found no changes in flux
density larger than our standard errors of about 5\% over the course of
an observing run, or, extrapolated, 10\% per day.
We therefore do not confirm the
occasional changes of up to 60\% (with inferred standard errors
of up to $\pm 30\%$) reported for similar short
periods by Ho \etal\ (1999).

\section {VLBI images \label{vlbmaps}}

\begin{figure}
\plottwo{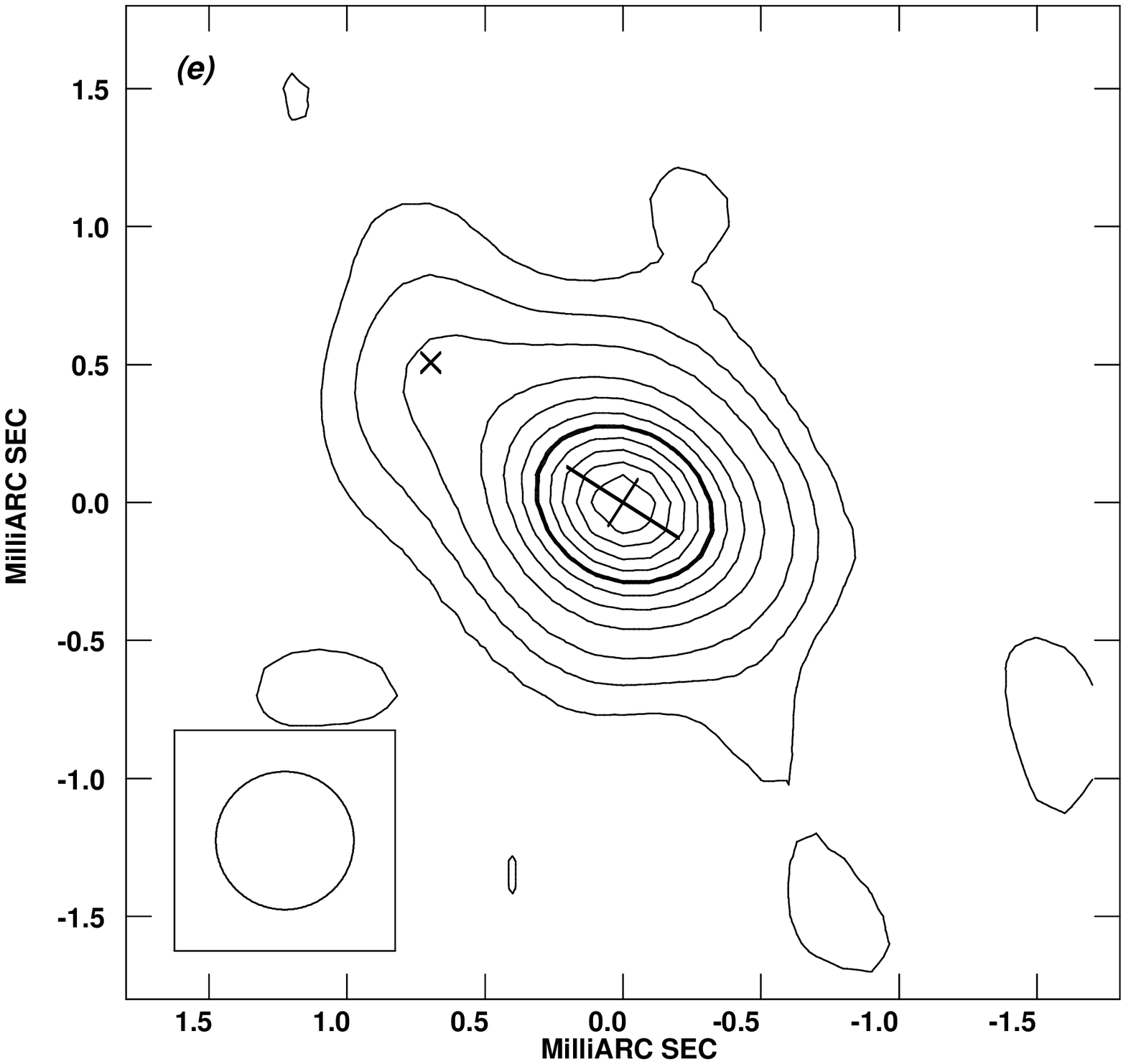}{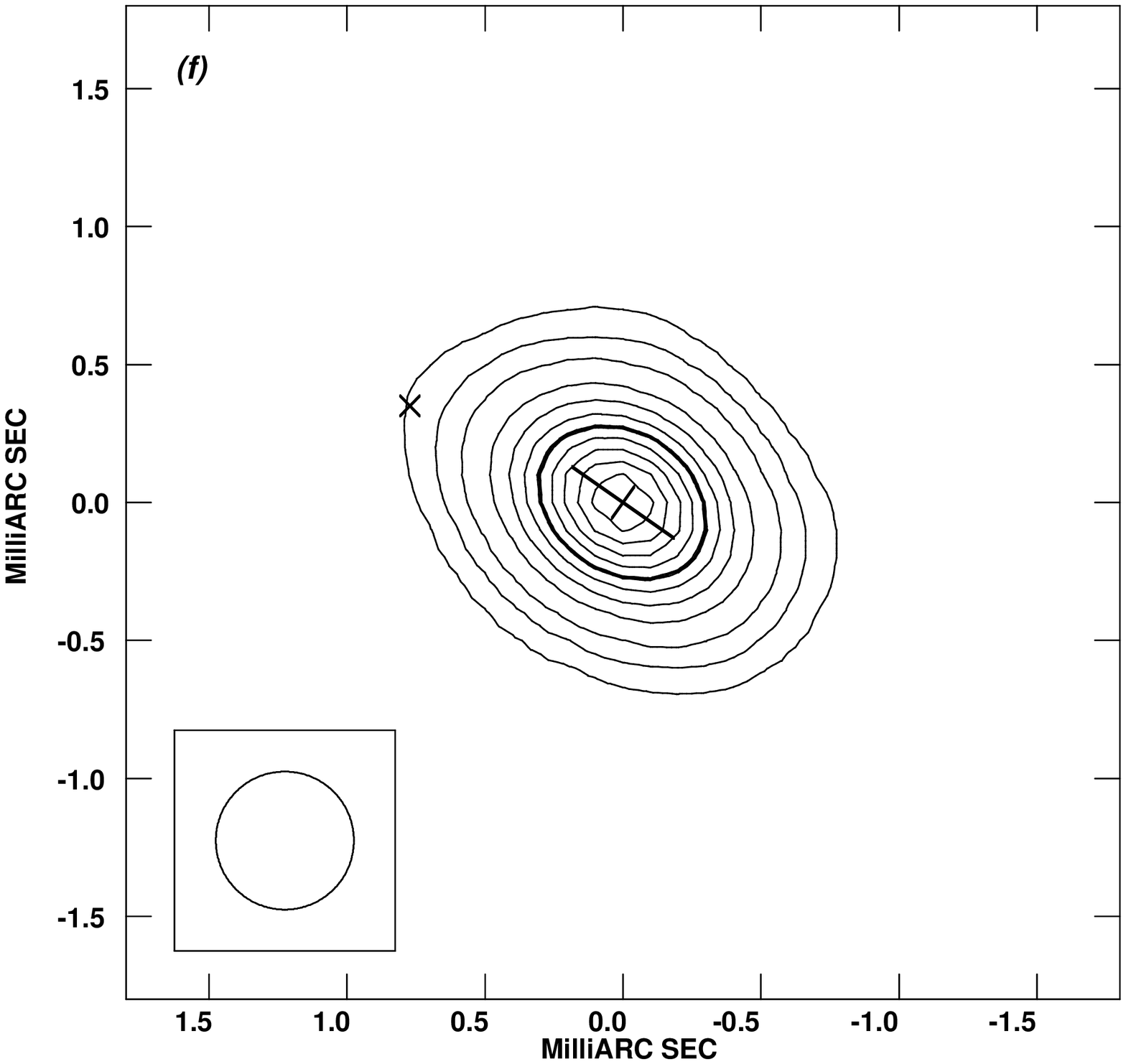}
\figcaption{CLEAN images of M81\* at 8.3~GHz.  The contours are at 2,
5, 10, 20, 30, 40, {\bf 50}, 60, 70, 80 and 90\% of the peak flux
density (the 50\% contour is indicated by a heavier line).  North is
up and east is to the left.  Uniform weighting was used, and the
images were convolved with a clean beam of FWHM 0.5~mas.  For three
images (see below) this beam represents a super-resolution by a factor
$\leq 1.8$.  However, because of the compactness of the source and the
relatively high dynamic range of $> 200$ our maps should be reliable
within the framework or our interpretations.  Also indicated are the
results of the two component model fit: the large cross shows the
position, orientation and FWHM size of the fit elliptical Gaussian,
and the small cross shows the position of the additional point source
that was fit to the data.  For each image, the observing date, the
peak brightness, the background rms, and the flux density of the
additional point source are given below.  Lastly, in parentheses, 
we give the uniform-weighted beam parameters (FWHM major and minor
axes, position angle) where 
its mean FWHM size was larger than 0.5~mas.\protect\\
$(a)$ May 17, 1993; 78~mJy/beam; $\pm 0.27$ mJy/beam; 6 mJy. \protect\\
$(b)$ June 21, 1994; 75~mJy/beam; $\pm 0.29$ mJy/beam; 18 mJy.\protect\\ 
$(c)$ Feb.~12, 1995; 101~mJy/beam; $\pm 0.21$ mJy/beam; 8 mJy
($0.90 \times 0.80$~mas at $-50\arcdeg$). \protect\\
$(d)$ May 11, 1995; 99~mJy/beam; $\pm 0.41$ mJy/beam;  8 mJy
($0.90 \times 0.60$~mas at $ 55\arcdeg$). \protect\\
$(e)$ Dec.~14, 1996; 90~mJy/beam; $\pm 0.28$ mJy/beam; 24 mJy
($0.80 \times 0.51$~mas at $-11\arcdeg$). \protect\\
$(f)$ Nov.~15 1997; 118~mJy/beam; $\pm 0.13$ mJy/beam; 5 mJy. 
\label{vlbimap}}
\end{figure}

Figure~\ref{vlbimap} $a$ -- $f$\/ show six representative images of
M81\*, from May~1993, June~1994, Feb.~1995, May~1995, Dec.~1996, and
Nov.~1997 respectively.  All the images are made with CLEAN, using
uniform weighting. They are shown convolved with a common restoring
beam of 0.5~mas FWHM (full width at half maximum). Even with the
resolution attained with a global array, the structure is only
marginally resolved.  Nevertheless, it is clear that the source is
extended at each of the six epochs approximately along the NE-SW axis,
with its length being of order 1~mas.

We searched for signs of extended emission on spatial
scales of $\sim 10$ to 100~mas, which are resolved out with VLBI but
unresolved with the VLA.  For all of our observing sessions, we
formed the ratio of the total flux density measured with the VLA to
that measured by VLBI.  This ratio for M81\* shows no systematic
difference to that observed for SN1993J, which physically is not
expected to have any extended emission on these scales.  We conclude
that the flux density of any emission on scales of $\sim10$ to
100~mas associated with M81\* is $<8$~mJy, or $<6\%$ of the total flux
density.

In addition to the extension of $\sim1$~mas, slight changes in the
structure are visible: the images shown in Fig.~\ref{vlbimap} show
that the size of the source along the NE-SW axis as well as the
orientation of its major axis are changing.  For example, both changes
in extension and orientation are evident between
Figs.~\ref{vlbimap}$a$ and $b$.

\begin{figure}
\plottwo{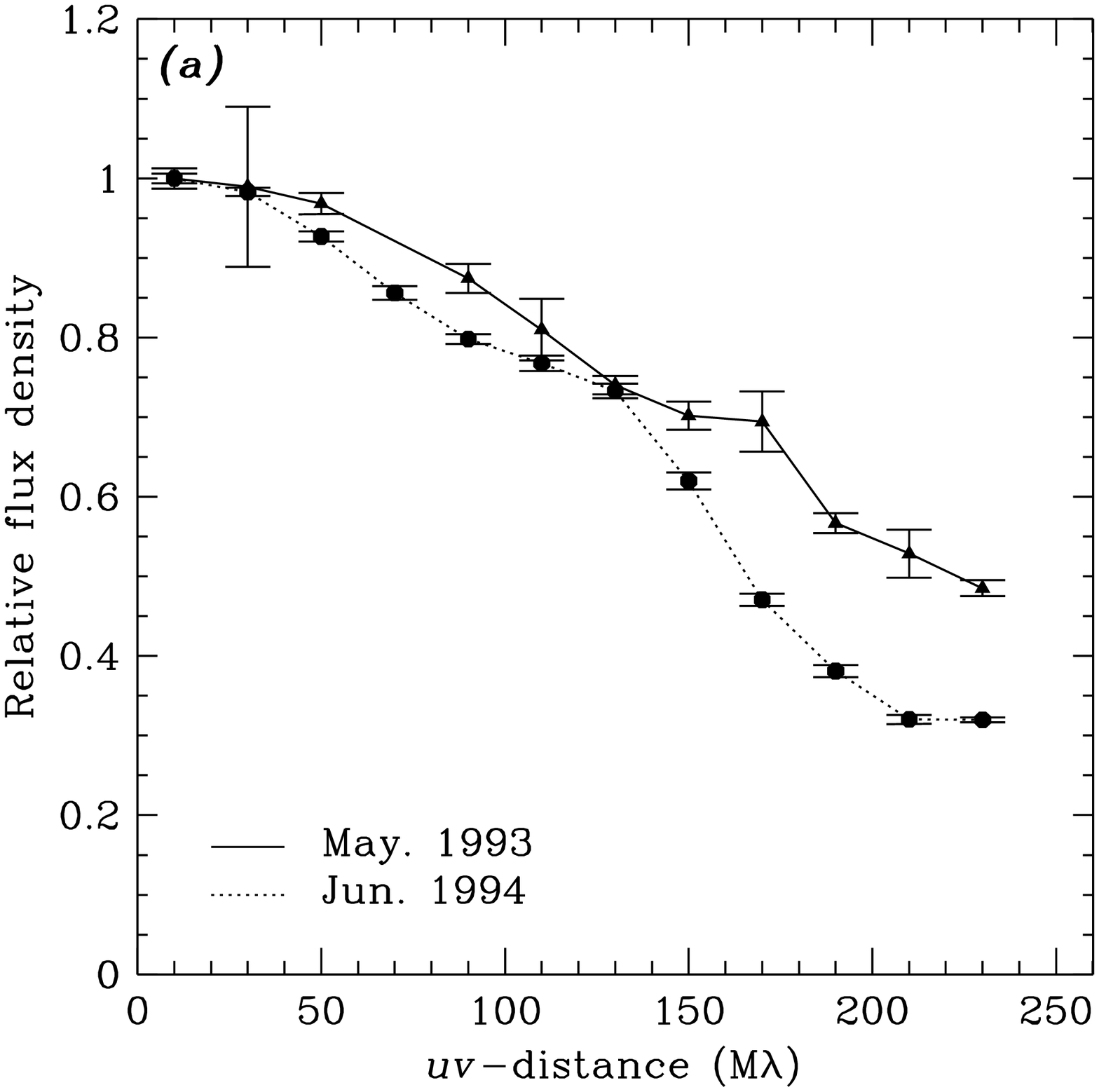}{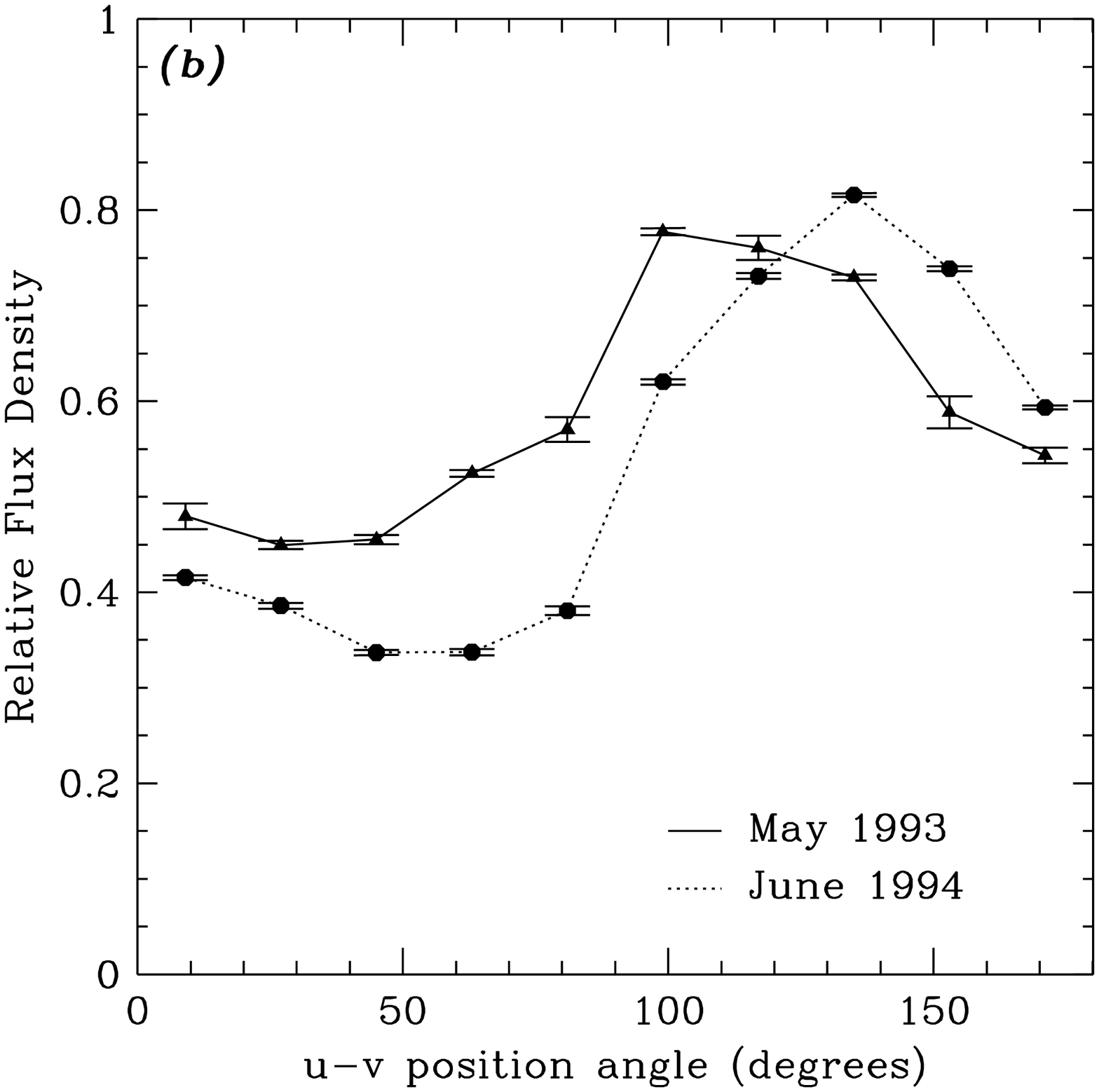}
\figcaption{Part $(a)$ shows the average visibility as a function of
\uv~distance.  The solid line denotes the May 1993 data and the dashed
line the June 1994 data, showing that the source is more extended in
June~1994.  Part $(b)$ shows, for the same two epochs, the average
visibility of M81\* in an annulus with \uv~distances between
200~M$\lambda$ and 260~M$\lambda$, as a function of \uv~position
angle.  The maximum and minimum of the two curves are clearly
displaced, indicating a shift in the p.a.\ of the major axis of the
source.
\label{uvvpa}}
\end{figure}

Are the variations significant given that the source size is generally
smaller than the minimum fringe spacing of the array?  Could they for
instance be artifacts of the self-calibration process?  To investigate
these questions, we searched for a signature of the different sizes
and orientations in the ``raw'' data, namely the \uv~data which have
only been self-calibrated with a preliminary model common to both
epochs, and were therefore still unaffected by the individual
iterative self-calibration processes for each epoch.  As an example of
the size change we show in Figure~\ref{uvvpa}$a$ the visibility
amplitudes as a function of the \uv~distance for our observing runs of
May 1993 and June 1994 (see Fig.~\ref{vlbimap}$a,b$ for the images).
Clearly the visibility curve for the data of the latter epoch is
narrower, showing that the source is indeed more extended at that
epoch.

Similarly, for the change in orientation, we show in
Fig.~\ref{uvvpa}$b$ the average visibility amplitude in an annulus of
\uv~distances between 200~M$\lambda$ and 260~M$\lambda$ as a function
of \uv\ position angle (p.a.) for the same two epochs. The limits of
the annulus were chosen to display the most significant variation of
amplitude with \uv~p.a.\ while excluding the noisier and less
frequently sampled outermost regions of the \uv~plane.  These curves
show a minimum and a pronounced maximum which indicate that the source
is not circular.  There is a clear displacement of one curve with
respect to the other of about 10 to 20\arcdeg\ of \uv~p.a.  Since the
minimum occurs at the p.a.\ of the major axis, the displacement
indicates different orientations of the source at these two epochs.
Furthermore, the minimum in the June 1994 data is deeper than in the
May 1993 data by about 30\% while the maxima have about the same
relative correlated flux densities.  This is a signature of differing
degrees of source ellipticity.

Since, again, these data have not been self-calibrated to convergence
for each individual epoch, the difference between the two curves in
each of the parts of Figure~\ref{uvvpa} cannot be ascribed to any
divergence in the self-calibration process, and therefore are highly
significant.  In fact, the self-calibration process generally caused
only small changes in our derived values, in particular, the plots
shown in Fig.~\ref{uvvpa} are virtually identical after
self-calibration.  We conclude that the size and orientation of the
source can vary significantly.  For the first time, structural changes
from one epoch to another have been shown in the ultracompact nucleus
of M81.

\section {VLBI model fitting \label{vlbmods}}
\subsection {One-component model and structural variability}

To investigate the structural variability more quantitatively we
employ modelfitting in the \uv~plane, which gives results which are
not dependent on convolution with a restoring beam.  First, we fit
only a single elliptical Gaussian to the data.  The results for 20
epochs are shown in Table~\ref{modfitt}.  Both the size and the p.a.\
of the Gaussian vary, by up to 0.46~mas (\ie\ 100\%) and 15\arcdeg\
respectively.
 
We determined the standard errors in the fit parameters as follows.
We divided an individual data set into three roughly equal parts in
time.  We repeated the model fit for each of these time periods, and
thus obtained three largely independent estimates of the model
parameters.  The standard error of each parameter is taken as the rms
of these three estimates scaled by $1/\sqrt{3}$.  Because of the small
number (ie.\ $n \! = \! 3$) statistics for each individual epoch, we quote
only an average standard error estimate derived from six
representative epochs.  In fact, we expect that this standard error
estimate represents, if anything, an over-estimate of the true
uncertainty because a disproportionately better fit, and thus smaller
errors, are possible when using all the data simultaneously.  We
consider this procedure slightly superior to that used in
Paper~I\footnote{In Paper I, we based our computation of the standard
error on the increase in $\chi^2$, which requires an estimate of the
number of degrees of freedom which is only approximately known.},    and
it yields consistent or slightly larger standard errors.

For each epoch the fit model parameter estimates and their standard
errors are listed in Table~\ref{modfitt}.  In addition the means of
the parameter estimates, their standard errors, and their rms values
are given at the bottom of the respective columns in the same table.

\subsection {Two-component model and structural variability\label{twocmod}}

\begin{figure}
\plotfiddle{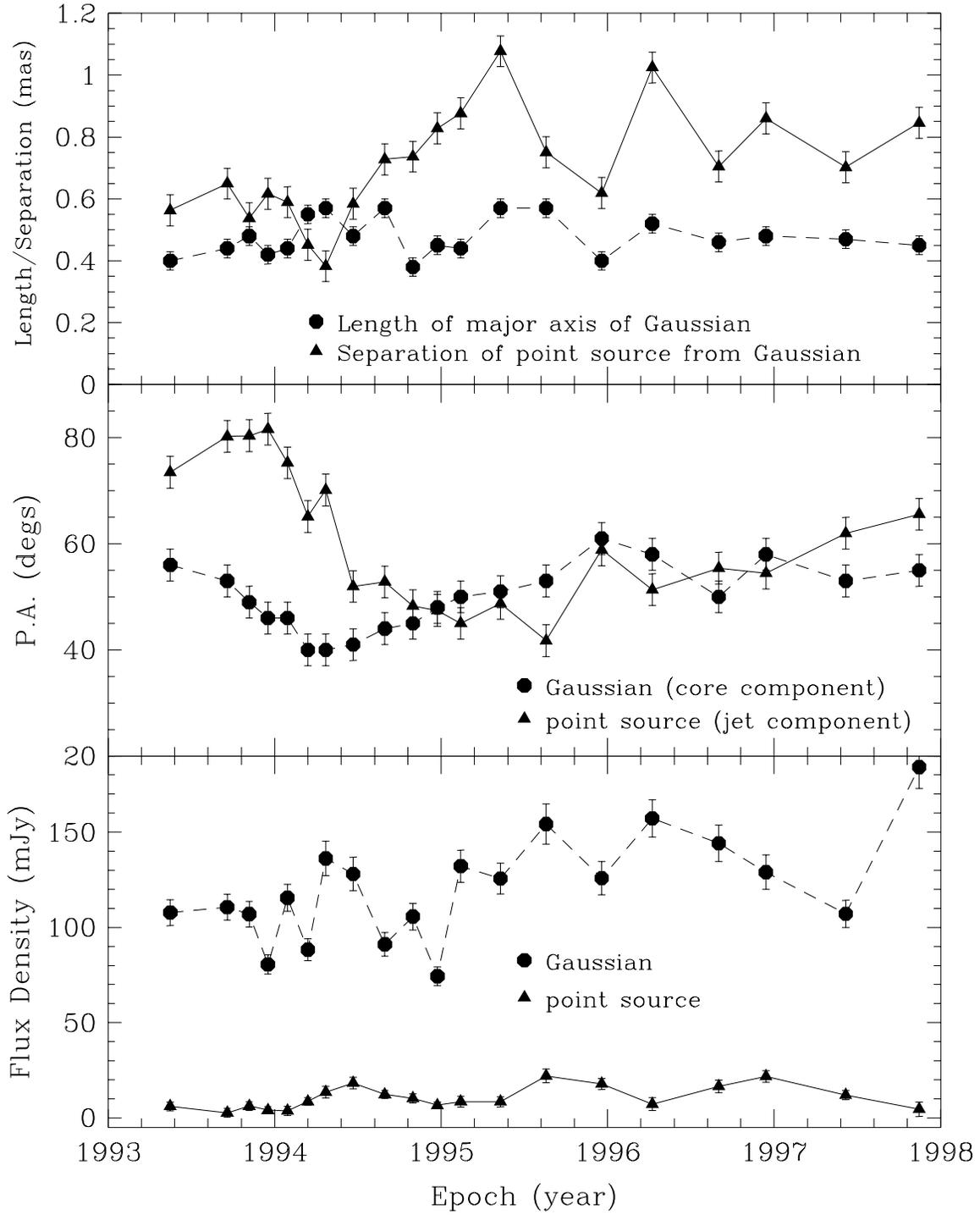}{7in}{0}{77}{77}{-230}{-40}
\figcaption{The variability of the parameters of M81\* as a function
of time. The parameters of the elliptical Gaussian are shown by
circles, while those of the additional point source are shown by
triangles.  The top panel shows the FWHM of the major axis of the
Gaussian and the separation of the point source from the Gaussian.
The central panel shows the p.a.\ of the major axis of the Gaussian
and the p.a.\ of the position of the point source with respect to the
Gaussian.  The lower panel shows the flux densities of the Gaussian
and the point source (normalized to the total flux density measured at
the VLA).  The uncertainties shown are typical standard errors which
include both statistical and systematic contributions (see text,
\S{\ref{vlbmods}}). \label{parplt}}
\end{figure}

Figure~\ref{vlbimap} shows that the source is slightly asymmetrical. A
single elliptical Gaussian therefore provides only an approximate fit
to the data.  We can quantify the asymmetry of the source by fitting a
point source in addition to the elliptical Gaussian; we use a point
source because it is the simplest enhancement to the Gaussian model.
In all cases we find that this addition causes a significant decrease
in $\chi^2_\nu$ (reduced $\chi^2$), typically by $\simgeq 10$\% (where
the number of degrees of freedom in the fit ranges from $\sim 500$ to
$10,000$).  Still more complex models did not improve the fit
sufficiently to warrant their consideration, given the limited
resolution of our observations. The standard errors of the fit
parameters were determined as described for the one-component model.
The results of these new fits are also listed in Table~\ref{modfitt}
and are plotted in Figure~\ref{parplt}.

At all epochs the elliptical Gaussian is the dominant component while
the point source has only 2 to 13\% of the total flux density.  The
point source is always located NE of the elliptical Gaussian.  Tests
in which we inverted the flux density ratio of the two components and
attempted to fit a faint elliptical Gaussian and a dominant point
source to the SW instead gave poorer fits, confirming our finding in
Paper~I that the source becomes fainter towards the NE.

Apart from the minor axis of the elliptical Gaussian, which remains
largely unresolved, all parameters vary by more than three times their
standard errors. The largest relative variations are found for the
parameters of the additional point source. The variations of the flux
density, separation from the elliptical Gaussian, and position angle
are all larger than the variations of the equivalent parameters of the
elliptical Gaussian.  Each of the three pairs is plotted for better
comparison in a panel of Figure~\ref{parplt}.

Of the six curves, those of the p.a.\ of the orientation of the
elliptical Gaussian and the p.a.\ of the position of the point source
with respect to the Gaussian (Fig.~\ref{parplt}$b$) appear to be the
smoothest. The former p.a.\ varies smoothly between about
40\arcdeg\ and 60\arcdeg\ and the latter between about 45\arcdeg\ and
80\arcdeg, both with a timescale of order one year.

The Gaussian and the point source are largely oriented along the NE-SW
axis, with the point source closely tracking the p.a.\ of the Gaussian
except during the period 1993-1994.  This suggests that the underlying
structure is mostly linear, as would be expected of a jet, but that it
is significantly bent during 1993.  Occasionally, apparently at times
when M81\* is compact, the p.a.\ can increase at least to 60\arcdeg\ for the
Gaussian and 80\arcdeg\ for the point source. The emission away from
the center therefore seems to be contained in a rather wide cone
between about 40\arcdeg\ and 80\arcdeg\ on the sky.

The variations for the separation of the point source from the
Gaussian (Fig.~\ref{parplt}$a$) and for the flux density of the point
source (Fig.~\ref{parplt}$c$) are somewhat less smooth. Finally, quite
erratic fluctuations appear for the remaining two parameters, the length
of the major axis and the flux density of the Gaussian
(Fig.~\ref{parplt}$a, c$), with timescales less than several weeks.

In order to investigate quantitatively how much the model parameters
evolve in a smooth rather than a random way we performed a non-parametric
test. 
For each of the six model parameters plotted against time in
Figure~\ref{parplt}, we computed rms$_f$, which is the
rms of the {\em changes}\/ of the model parameter values from each epoch
to the following one. If the model parameter values were randomly
distributed over time, these changes would show no correlation with their
respectively succeeding values. Changing the time sequence of the values
would therefore make no difference to the rms$_f$ value. By contrast,
if the model parameter values displayed a smooth evolution, then the
rms$_f$ value for the observed sequence would be smaller than that for
a random reordering of the values.

Based on 5,000 random reorderings of the values of each of the six
parameters we found that the probability $p$ of obtaining rms$_f$
values as low as those observed for the p.a.s of the Gaussian and of
the position of the point source is $< 0.1$\%, confirming that these
two parameters are indeed varying smoothly and not randomly from epoch
to epoch.  For the separation of the point source from the Gaussian
and for the flux density of the point source the same statistic, $p$,
is $\sim3\%$, confirming that these parameters vary less smoothly.
For the remaining parameters, the length of the major axis and the
flux density of the Gaussian, $p \sim 50$\% and 10\%, respectively,
confirming the more erratic nature of their variations.

The smooth evolution of some of the model parameters strongly suggests
a smooth evolution of the source structure on similar timescales.
While
our \uv~coverage varied from run to run, it varied in a
non-systematic way (see Table~\ref{antab}) which would be very
unlikely to cause the smooth variation seen in some of the model
parameters.  Further, though the true source structure is undoubtedly more
complex than our simple model, the smooth changes in the model on
a timescale of $\sim 1$~year would be very unlikely unless the true source
geometry were also changing smoothly on the same timescale.  In
other words, not only have we detected significant temporal changes in
the source structure, but, more specifically, temporal changes that
are characterized by a smooth evolution in case of the orientation of
the Gaussian and the point source, a somewhat less smooth evolution in
case of the distance of the point source from the Gaussian, and in
contrast, completely erratic fluctuations in case of the total flux
density of the source.  These characteristics are consistent with the
behavior expected from a core-jet source, which we have already
suggested in Paper~I as the underlying structure in M81\*

\section {The core \label{core}}
\subsection {Where is the core? \label{coreposs}}

If the nucleus of M81 does have a core-jet structure, then where is
the core located?  Is the core to be identified with the point source,
or does the core lie at the center, or perhaps near one end of the
Gaussian?  The most direct and reliable method of identifying the core
is to determine astrometrically the motion of components in a core-jet
source with respect to a ``reputable'' reference source. The core,
being associated with a supermassive black hole, must remain (largely)
stationary, while the jet may move (see Bartel \etal\ 1986 for
such an identification of a core). For the possible core-jet source
in M81 we used the geometric center of SN1993J as a reference
point. Any motion between the core, taken to be the gravitational
center of the galaxy, and the geometric center of SN1993J would be
constrained to (1) the orbital motion of the progenitor around the
gravitational center of the galaxy, (2) any peculiar motion the
progenitor might have had, and (3) any motion of the center of SN1993J
resulting from an asymmetry in the explosion.  The velocity of (1) can
be estimated independently from the HI rotation (Rots \& Shane 1975)
to be $250 \pm 80$~\kms\ at p.a.~$-40\arcdeg$.  The velocity of (2)
can be expected to be $\simleq 100$~\kms.  And, the velocity of (3),
judging from the high degree of circular symmetry of the supernova
images (to within 3\%, Bartel \etal\ 1999), can be expected to be
$\simleq 500$~\kms. In all, after subtracting the orbital motion (1),
we would not expect the velocity difference between the core and the
center of SN1993J to be larger than about 500~\kms.
 
The technique of phase-referenced mapping allowed us to determine the
position of M81\* with respect to the geometric center of SN1993J with
high accuracy.  SN1993J is located in the galaxy M81 but is physically
unrelated to M81\*, and therefore it can, for our purposes, define a
galactic reference frame.  Since the two sources are separated on the
sky by only $\sim 0.3$\arcdeg, the dominant sources of error in VLBI
measurements of relative positions are much reduced.  In particular,
we estimate that the combined effects of statistical uncertainty and
errors in the {\em a~priori}\/ coordinates of M81\*, antenna
coordinates, tropospheric and ionospheric delays at each station, UT1,
pole position, and those caused by the variable structure of SN1993J
and M81\* give a standard error of $\sim 50$~\muas\ for each
coordinate and for each of the epochs (see, e.g.\
Marcaide and Shapiro 1984, Bartel \etal\ 1986, Rioja \etal\ 1997 for
astrometric observations with similarly small standard errors).

\begin{figure}
\plotfiddle{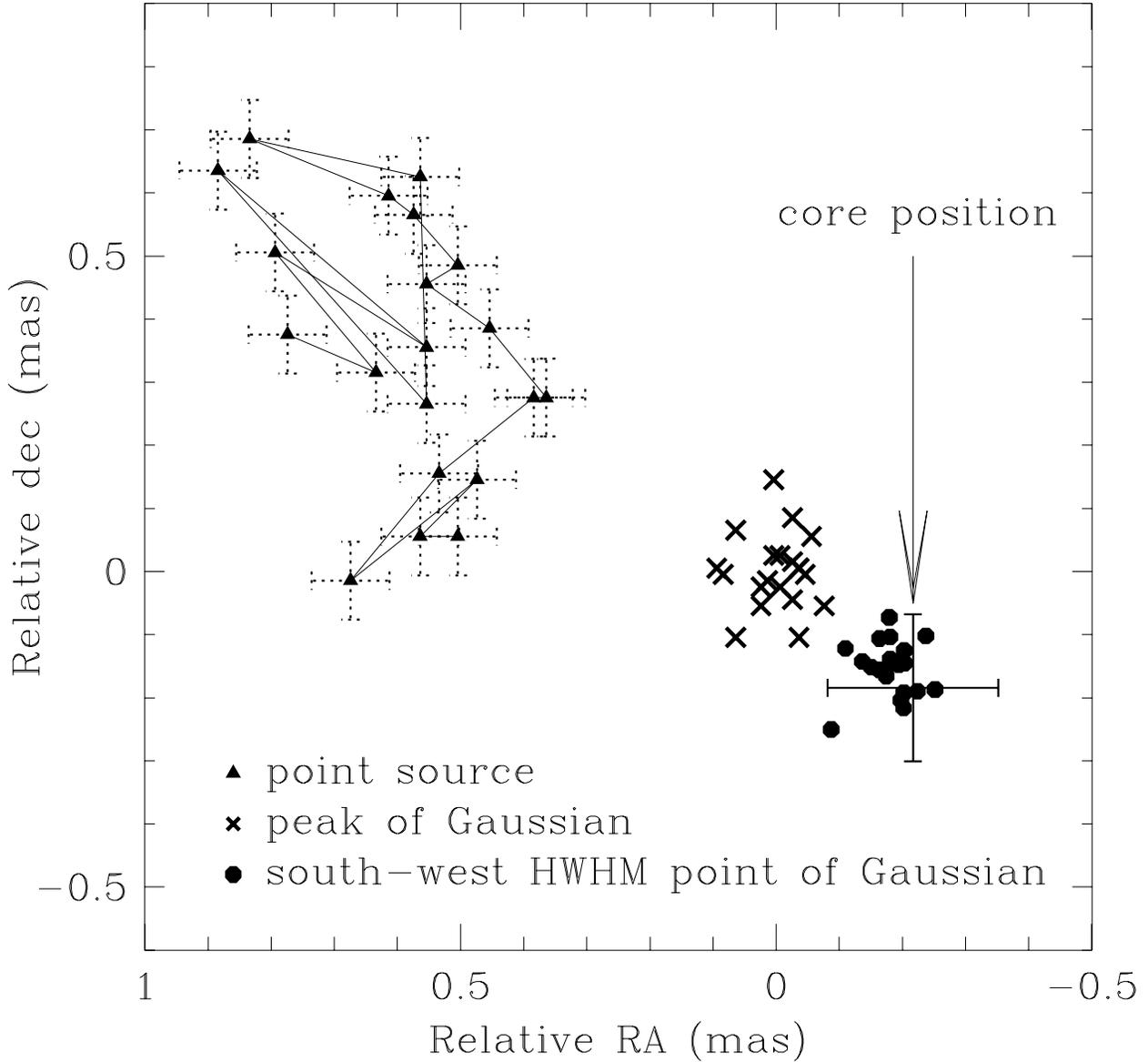}{6in}{0}{83}{83}{-260}{-30}
\figcaption{Positions at each epoch of three points in the two
component model, plotted with respect to the geometric center of
SN1993J.  The three points are: $i)$ the point source, shown by
triangles, with the motion indicated by the connecting line where the
lower right end of the line is the position in May, 1993; $ii)$ the
center position of the Gaussian, shown by the small crosses (with the
standard errors of 50~\muas\ omitted for clarity); and $iii)$ the
positions of the south-west  HWHM point of the Gaussian, shown by the
circles (again with the standard errors of 60~\muas\ omitted for
clarity).  The origin is the mean position of the center of the
Gaussian with respect to SN1993J.  Also shown is the derived core
position with its standard error.
\label{astrom}}
\end{figure}

In Figure~\ref{astrom} we plot the positions relative to the
geometric center of SN1993J of three particular points in the
two-component model for each of the 20 epochs.  These three points
are, from the NE to the SW, the point source, the center of the
dominant Gaussian\footnote{We actually plot the position of the phase
center, as determined by a single Gaussian fit.  The position of the
dominant Gaussian in the two component fit is shifted by a small
amount from this phase center.  Tests show that this shift is $20 \pm
20$~\muas\ in each RA and Dec., and our positional standard errors
include this contribution.}  and the SW half-width at half-maximum
(HWHM) point of the Gaussian.  It is apparent that the scatter of the
positions for these three points decreases towards the SW. The rms of
the positions for the point source is $600 \mu$as, that for the center
of the Gaussian is $80 \mu$as, and that for the SW HWHM point is $60
\mu$as.

\begin{figure}
\plotfiddle{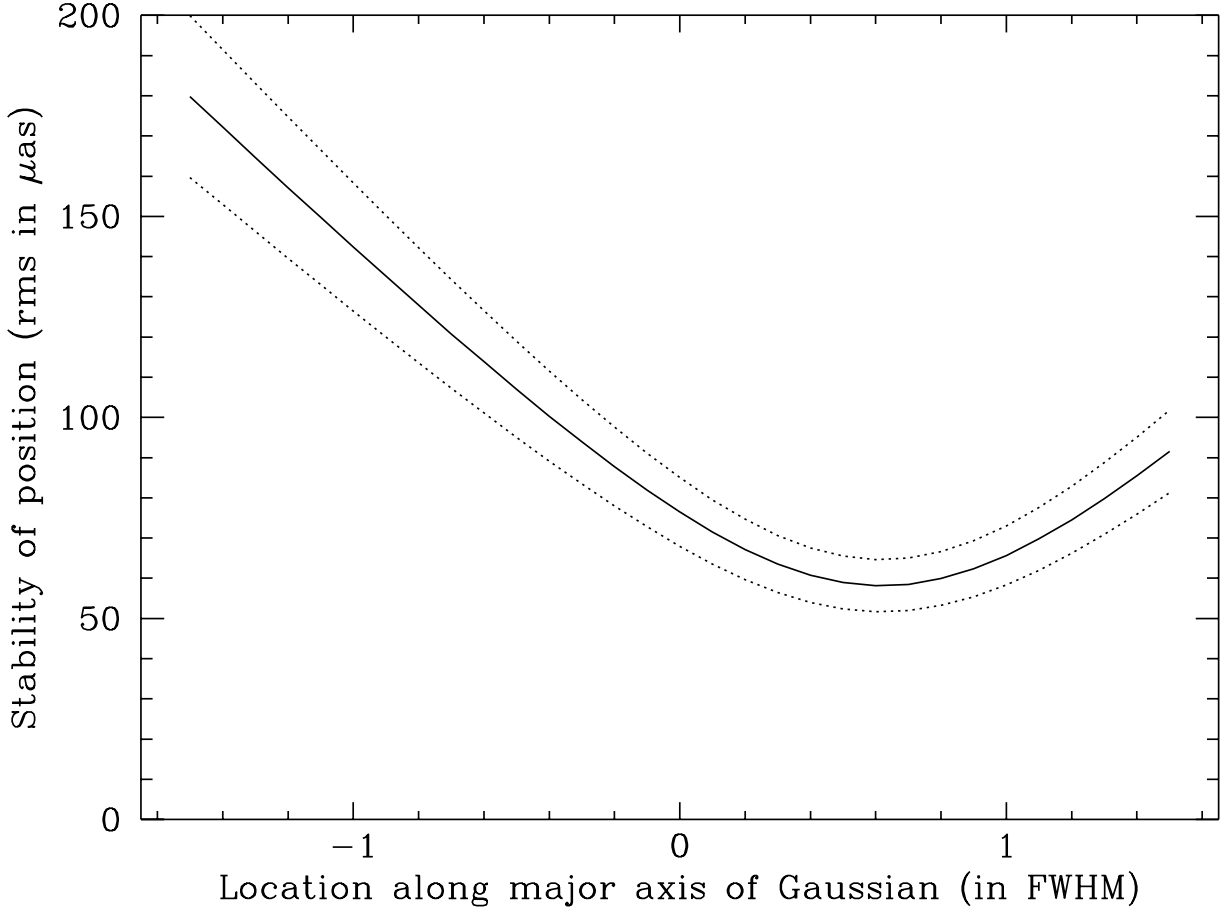}{5in}{-90}{63}{63}{-260}{390}
\figcaption{The solid line shows the variability in position of
different locations along the major axis of the fit Gaussian.  The
variability is the rms in \muas\ of the position on the sky of that
particular location on the major axis of the Gaussian.  The minimum
variability indicates the position of the core along the major axis of
the Gaussian.  The standard errors of the rms variability are indicated by
the dotted lines, from which the uncertainty of the core position is
derived.  The curve is smooth and continuous because the position,
and thus the positional rms over 20 epochs, is functionally defined
for any point along the major axis of the Gaussian.
\label{coreposf}}
\end{figure}

The gradient of the rms values along the major axis of the source
indicates that the astrometrically most stable point is located close
to the SW end of the Gaussian. In fact, we can locate the stable point
more precisely: in Figure~\ref{coreposf} we plot the positional rms as
a function of location along the major axis of the Gaussian.  The
minimum in the curve indicates the most stable point in the model
geometry. That point is located at $0.6 \pm 0.3$~FWHM from the center
of the Gaussian towards the SW, \ie\ close to the 35\% contour of the
Gaussian.  The rms of the positions of this point on the sky is only
58~\muas, approximately equal to the observational standard error
of our astrometric position determinations of 50~\muas\ mentioned above.

We have thus determined the most stationary point in our geometrical
model of the brightness distribution.  Is there any meaning to be
associated with this point or could it be that the more complex true
source geometry happens by chance to be least variable at this
particular location?  We think the latter to be unlikely, since the
variability in position not only reaches a minimum at that point, but
is consistent with being zero, which would be unlikely to occur by
chance.  Thus, we can conclude that we have identified a real,
stationary part of the brightness distribution.

Can this stationary point be identified with the core?  Since the core
is expected to be a compact component, it might be asked why we do not
parameterize the source with a compact component representing the
core.  However, no particular core component is apparent in the
brightness distribution at any of the 20 epochs, and, as mentioned in
\S\ref{twocmod}, when we attempted to fit a model consisting of a
point source and an elliptical Gaussian to the NW of it, we obtained poorer
fits.

Therefore, any core component that may be hidden in the brightness
distribution has to be rather weak. It follows that the brightness
distribution we see is dominated by emission from the jet.  It
then also follows that it is emission from the jet, which mostly
determines the size and location of the model components which we use
to parameterize the brightness distribution, and thus which mostly
determines the location of our stationary point.
This point is either the core itself, or the turn-on point of the jet.
Even in the latter case, we consider it physically unlikely that
the turn-on point is at any significant distance from the core.
We therefore identify the stationary point with the
{\em location}\/ of the core with considerable confidence.  

We plot the location of the core in Figure~\ref{astrom}.  Note that
the uncertainty in the core position is dominated by the uncertainty
of locating the stationary point within our model, rather than by
measurement uncertainties of our positions.  It is remarkable that,
despite the simple structure of our geometric model, we can identify a
point in it which is stationary --- this result suggests that our
model describes the source geometry reasonably well even for a
resolution significantly smaller than that obtained for the images.
The location of the core within the brightness distribution allows us
to determine that $\simleq 25$\% of the emission comes from the core,
and $\simgeq 75\%$ from the jet, whose properties we will discuss
further in \S\ref{jet} below.

To summarize, we have found a point in the brightness distribution,
which, to approximately within the observational standard errors, is
at rest in the galactic reference frame. We therefore identified
this point, with an uncertainty of 600~AU, with the location of the
core.

\subsection {Upper bound on the proper motion of the core \label{coreprop}}

Having argued in \S\ref{coreposs} above that the relative velocity
between the stationary core and the center of SN1993J is expected to
be $\simleq 500$~\kms, what in fact do we measure?  For the four and a
half year period the angular velocity is $9 \pm 7$~\muas~yr$^{-1}$ in
R.A. and $10 \pm 7$~\muas~yr$^{-1}$ in Dec.  At the distance of M81,
this corresponds to $160 \pm 110$ and $170 \pm 120$~\kms\
respectively.  Subtracting the expected galactic motion of SN1993J of
250~\kms\ at p.a.~$-40$\arcdeg\ (\S~\ref{coreposs}; calculated from
the HI rotation curve) and the inclination angle of the galaxy
(59\arcdeg), leaves a peculiar velocity of the core of M81\* relative
to SN1993J of $360 \pm 180$~\kms. This value is indeed comparable to
the peculiar velocity expected for SN1993J, which we discussed earlier
(\S\ref{coreposs}).  We do not, however, consider this value to be
significant.

\section {The jet  \label{jet}}

\begin{figure}
\plotfiddle{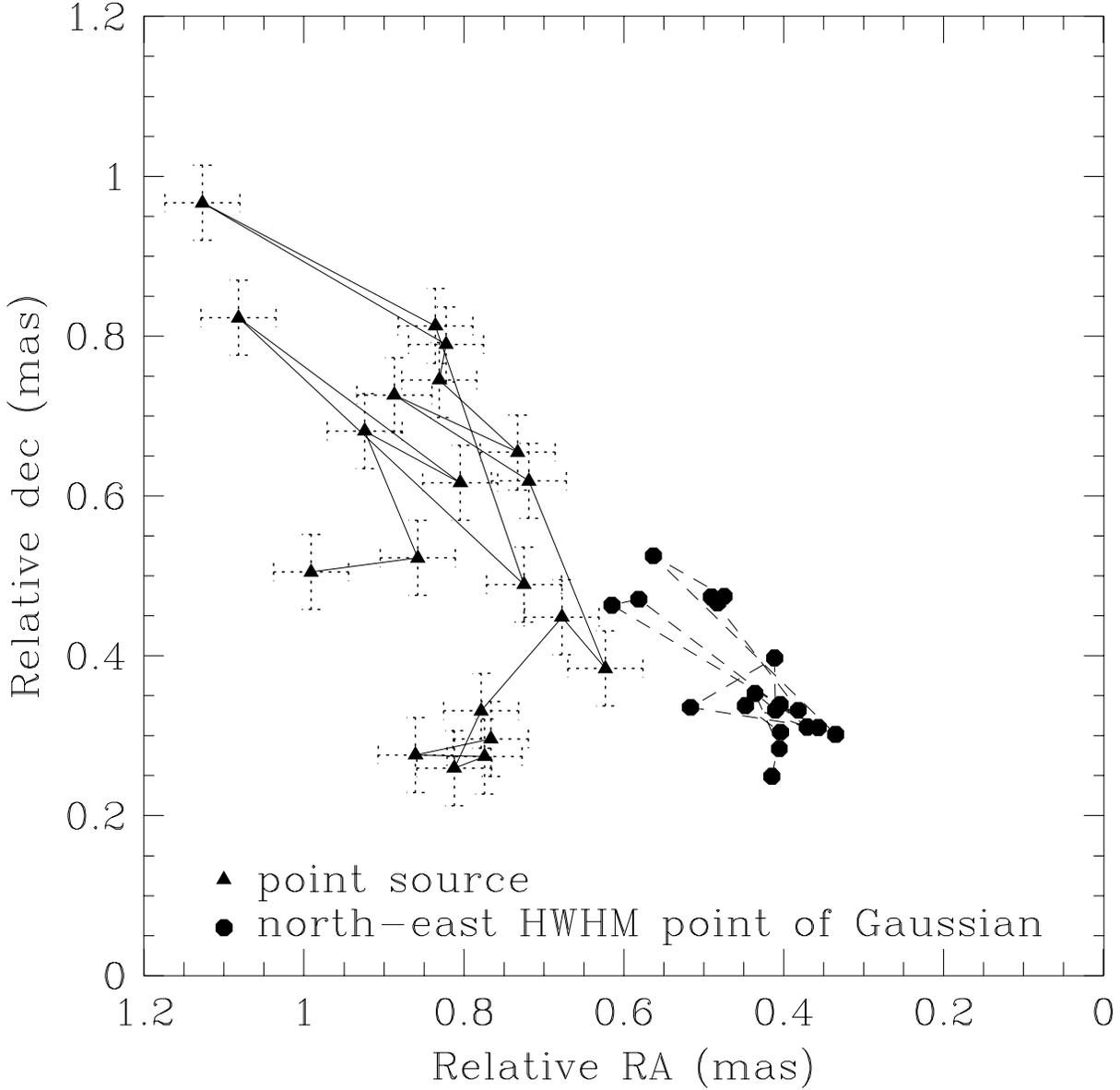}{6in}{0}{83}{83}{-260}{-30}
\figcaption{Positions at each epoch of two points in the two component
model, plotted with respect to the core position.  The triangles
(connected by a solid line) show the positions of the additional point
source, with the lower right end of the line indicating the position
in May 1993, and the upper left end corresponding to the position in
Nov.~1997.  The circles, connected by a dashed line, show the
positions of the north-east HWHM point of the Gaussian, \ie\ that away
from the core, over the same time period (with the standard errors of
30~\muas\ omitted for clarity).
\label{jetfig}}
\end{figure}

We have shown that the bulk of the emission originates in the jet, and
we now proceed to discuss its characteristics.  It is predominately
oriented at a p.a.\ of 50\arcdeg, and occasionally bends towards the
east.  The point source represents the distant end of it. How does the
jet move with respect to the core?  In Figure~\ref{jetfig} we plot the
positions with respect to the core of two points in the jet for each
epoch: the NE HWHM point of the dominant Gaussian, which represents
the inner jet, and the point source, which represents the end of the
jet. This figure is similar to Figure~\ref{astrom}, but the reference
point is now taken to be the core position, rather than the geometric
center of SN1993J.  Again, it can clearly be seen that the inner jet
as well as the end of the jet move, over a range of 0.3~mas and 0.8
mas, respectively.  The motion is predominately along the major axis
of the source.

\subsection {Slow motion of the jet with $v < 0.08 c$}
 
We searched for signs of a continuous slow motion in the source by
fitting a straight line to the length of the major axis as a function
of time.  Neither in the case of the single component model nor in
the case of the two-component model is the linear term significantly
different from zero.  The nominal values for the change of the 
major axis length with time are $-3 \pm 15$~\muasyr\ in R.A. and $4 \pm
9$~\muasyr\ in Dec., corresponding to $-50 \pm 250$~\kms\ and $70 \pm
160$~\kms\ respectively.

Is there coherent motion over shorter periods?  Only during two
periods early on, when our sampling was more frequent, did the
variation in length appear smooth. The corresponding longitudinal
speeds of the end of the jet (\ie\ the point source) are close to zero
during the first period from 1993 to 1994 and to 0.7~\masyr\
(12,000~\kms, $0.04 c$) during the second period from 1994 to 1995.
This latter motion corresponds in Figure~\ref{jetfig} to the long
connecting line from a distance of 0.4~mas to 1.1~mas from the
center. Three representative images of M81\* during this period are
displayed in Figure~\ref{vlbimap}$b - d$. The lengthening of the jet
is obvious.  For later epochs the time interval between observations
was longer and likely prevented us from temporally resolving the
motion. The fastest length changes occurred between Dec.\ 1995 and
Apr.\ 1996 and correspond to an outward motion of $0.075 c$.

It is interesting that the transverse velocities for the end of the
jet, implied by the changing p.a.\ and the (projected) length of the
jet, are of the same order of magnitude as the longitudinal speeds
during the period (\ie\ 1993.5 to 1994.5) when the p.a.\ of the point
source is changing.  If we assume that the temporally unresolved
motions of the end of the jet at later times are not significantly
higher than the temporally resolved motions at earlier times, we find
that the end of the jet moves in any direction on the sky with
velocities of up to $0.08 c$.  These velocities are large in
comparison to the velocity bound of the core, but small in comparison
to relativistic velocities commonly observed for jet components.

\subsection {Fast motion in the jet with $v > 0.25 c$}

It is remarkable that during large total flux density changes there is
comparably little, if any, change in the structure. In particular for
relatively large flux density increases over short time intervals,
e.g.\ 50\% between 15 March and 22 April 1994, the length and
orientation of the single as well as the two-component Gaussian model
do not change significantly.  The increase in brightness must be
spread out equally over most of the area represented by the dominant
Gaussian.  The speed must then be at least that given by the
effective size of the Gaussian divided by the time interval between
the epochs.  Tests with a model show that the flux density needs to
spread over $> 75\%$ of the FWHM of the Gaussian in order for the
apparent change in the size of the Gaussian to be less than our
standard errors of $\sim 10\%$. Therefore our lower limit for the speed
with which the flux density is transported away from the core along
the jet is 0.4~mas between the above two dates, which is equivalent to
$0.25 c$.

Given that the FWHM length varies quasi-randomly from epoch
to epoch, with an rms of $\sim 20$\%, the lack of a significant change
of the length could perhaps be a chance coincidence.  The next largest flux
density increases over short time intervals are 40\% between 17 Dec.\
1993 and 29 Jan.\ 1994 and 75\% between 23 Dec.\ 1994 and 12 Feb.\
1995, and again no significant change in the length of either of the
two models was found.  The fact that none of the sudden increases in
flux density in our data are accompanied by a significant change in
size suggests that our lower limit on the speed is realistic.

Even more rapid variations in flux density are reported by Ho \etal\
(1999).  If they also occurred without any accompanying structural
changes then an even faster transportation speed in the jet would have
to be inferred.  The fastest variation they report, $60 \pm 30\%$ per
day, would be equivalent to a lower bound on the speed of $4 c$.
However, as mentioned in \S\ref{fluxds}, our observations do not 
confirm such large intraday variations.  

We also searched our VLBI data for direct signs of extremely rapid
outward motion within an individual observing run. We monitored the
position of the point source on timescales of several hours by
splitting datasets into several segments and using model fits for each
segment, but no coherent outward motion was found, giving a $2\sigma$
upper bound for any such motion of $14 c$.

\section{Discussion \label{discuss}}

Twenty epochs of observations of the nucleus of M81 phase-referenced
to a physically unrelated source in the same galaxy have significantly
advanced our knowledge of this galaxy's center.
At a distance of only 3.63~Mpc, M81 is the nearest spiral with an AGN
and therefore of particular interest for the study of our own Galaxy's
central source.  In fact M81 is, besides Cen~A, the nearest galaxy with
an AGN altogether.  But despite its closeness the nucleus can be
only partly resolved even with the longest interferometer baselines on
earth.

Previously we reported on the basic size, elongation, and orientation
of the source structure and their dependence on frequency as well as
on the evidence that the brightness decreases towards the NE (Paper I,
B82). The source structure remained remarkably unchanged from one
observation in 1981 to the start of our observations in 1993. We
tentatively interpreted the results in terms of a core with a bent
jet.  The present observations draw a clearer picture of the nuclear
source.  Phase-referenced mapping enabled us to identify, with a
standard error of only 600~AU, a stationary point.  This point, located
SW of the peak of the brightness distribution, is most likely the
location of the core. The velocity within our galactic frame of
reference is only $\leq 730$~\kms ($2\sigma$ upper bound).

The uniqueness within the galaxy of M81\* and the relatively low velocity
limit of the stationary point in it suggests that it is associated
with the putative black hole in the gravitational center of the
galaxy.  In our Galaxy, stars with velocities as high as $\sim
1,000$~\kms\ have been found within 0.01~pc of Sgr~A\* (Genzel \etal\
1997). Only Sgr~A\* itself appears to be at rest, having a proper
motion with respect to extragalactic sources of $\simleq 20~$\kms\
(Reid \etal\ 1999), which is consistent with the expectation for a
supermassive black hole in the center of our Galaxy.  Similarly, M81\*
having a stationary point in its structure is consistent with the expectation for
such a black hole in the center of M81.

This identification of the core is important for further studies of
the core-jet system of M81\*, for instance to determine
unambiguously the spectrum of the core and the spectral index
distribution of the jet (for a preliminary discussion see Ebbers
\etal\ 1998).

The identification of the core is also important for studies of
SN1993J.  We have now found a stable reference point within M81\* ---
the gravitational center of the galaxy --- with which we can determine
with the highest accuracy possible the position of the explosion
center of SN1993J and investigate the expansion of the supernova with
respect to it.

Furthermore, the identification of the core will become important for
efforts to determine the proper motion of the galaxy M81 in the
extragalactic reference frame, which would have implications for
models of the dynamics of nearby galaxies.  No proper motion of any
galaxy has yet been measured, but the stationary core in the
gravitational center of M81 is the ideal source to measure such a
proper motion given a sufficient time baseline of $\simgeq 20$~years.

The remaining part of the structure varies in length and orientation
about the core, and is therefore identified with the jet. The
orientation of the jet displays the smoothest variations, with a
timescale of $\sim 1$~yr. Occasionally the jet bends by about
20\arcdeg\ towards the east.  This bending is consistent with the
previously reported increase in size and p.a.\ of the source towards
lower frequencies.  The length of the jet varies less smoothly, with
the end of the jet moving longitudinally with (projected) velocities
of $< 0.08 c$. The end of the jet also moves transversely, with similar
velocities.

The most erratic variations occurred for the flux density of the
dominant Gaussian (and the total flux density of M81\*, which is
dominated by the latter). Refractive interstellar scintillation can be
excluded as the cause of the variations, since for a source size of
say 0.3~mas it could account for variations of at most a few percent
at 8.3~GHz (Rickett 1986).  Our flux density variations are much
larger and thus clearly intrinsic to the source.  We can further
conclude that the structural variations are also intrinsic to the
source, since they are equivalent to local brightness changes of more
than a few percent, and since they are non-random from epoch to epoch.

The hypothesis that the energy source for galactic nuclei is a
starburst has been advanced by Terlevich \etal\ (1992).  We showed in
Paper~I that such a hypothesis is untenable for the center of M81\*
because of its compactness and low average expansion velocity.
Recently, Wrobel (1999) has shown that the starburst hypothesis is
unlikely for another low level AGN, NGC5548, on the grounds of the
astrometric stability of the radio nucleus.  We note that in this
paper we have shown that the motion of M81\* is more than 20 times
less than the upper limits for NGC5548 measured by Wrobel (1999), and
this only strengthens our conclusion from Paper~I that the starburst
hypothesis is not applicable to M81\*.

How do these results fit into the framework of the core-jet models of
M81\* suggested by previous observations?  In early investigations
M81\* was modelled as an optically thick synchrotron self-absorbed
source. It was interpreted as consisting of either one component
elongated approximately along the rotation axis of the galaxy with
magnetic field and electron energy density distribution declining
along it, or as a few homogeneous condensations along the same axis,
each becoming optically thick at a different frequency (B82, see also
de Bruyn \etal\ 1976).

In Paper~I we concluded that a bent, steep-spectrum jet probably
oriented towards the NE was the most natural interpretation of the
VLBI observations. We applied to M81\* a model of Sgr~A\* developed 
by Falcke, Malkan \& Biermann (1995) on the basis of Blandford \&
K\"{o}nigl's (1979) model of an optically thin unconfined jet.  This
allowed us to infer that the inclination angle of the jet is
$\geq 30\arcdeg$, and the accretion rate is $\sim
10^{-5.5}$~\Msol~yr$^{-1}$. 

Both Falcke (1996) and B\"{o}ttcher, Reuter
\& Lesch (1997) have since modelled the VLBI data for M81\* in more
detail.  In their models, the source consists of a combination of jet
condensations, which fade as they move outward and expand freely.
The jet condensations are expected to move relativistically,
consistent with our lower limit on the transport speed of $0.25
c$. The overall source size is expected to increase with frequency
$\propto \nu^{-0.8}$ as observed in Paper~I. The lifetime of the
condensations is short: applying the same parameters as B\"{o}ttcher
\etal\ (1997), we calculate from the present results that it is only
about two weeks.  As the time between our observations is longer than
two weeks, we expect to see a completely different set of
condensations at each epoch, and thus we would not expect to detect
any direct outward motion of jet condensations. Instead we would expect
rapid variation in the length of the jet, which we indeed observe. 
The total flux density, \ie\ the sum of the
flux densities of the individual condensations, would be expected to
vary on a timescale of the lifetime of the condensations --- again,
about two weeks.  This is also consistent with both our observations
and those of others (Crane \etal\ 1976; Ho \etal\ 1999).

In contrast, the smoother variations observed for the orientation of
the jet and occasionally for its length suggest changes which are
distinct from the fast flow of jet condensations or the plasma. They
could for instance be caused by precession of the central engine. It
is also possible that they reflect the changing orientation and
perhaps geometry of a variable channel or tube along which the
ejected condensations are traveling. In other words, the plasma flows
relativistically through a more slowly changing channel.

A possible geometric scenario is that the source behaves somewhat like
a garden hose, fixed at one end, but free to move at the other.  The
fixed end of the hose represents the end of the jet near the core.
The plasma flows through the hose with a high velocity, \ie\ $>0.25
c$, and becomes fainter as it moves outward.  The rate of injection at
the core is rapidly variable, and this rapid variation will cause the
observed total brightness and likely the illuminated length to vary
erratically.  In addition the orientation of the garden hose
itself varies, possibly driven by the rapid variability of the flow.
However this variation occurs more slowly, and this slower variation
causes the observed smooth changes of the orientation of the jet.  It
also causes smooth changes of the length of the jet as a projection
effect, since the variation of the orientation of the garden hose is
unlikely to be confined to the plane of the sky, and so a change in the
orientation would in general also cause the projected length of the
garden hose to vary.  Thus in addition to any rapid variations in length
caused by the rapid flow variability, there would be smoother
variations in projected length caused by the changing orientation of
the hose.  It is these last variations which would account for the
observed episodes of smooth evolution in length.

With the core-jet scenario now firmly established for M81\*, we can
see general similarities to the more powerful core-jet systems in
other AGN.  The latter are mostly one-sided, as is M81\*.  In fact,
only very few AGN have been shown to have counter-jets. 
A counter-jet in M81\* cannot be excluded particularly because some
emission has been detected with the VLA in the SW at a distance of
$\sim 0.1$~pc from the core (Bartel \etal\ 1995).
However, taking the uncertainty of the core position into account, any
VLBI counter-jet in M81\* would have a flux density of $\simleq 20$\%
of that of the remaining source at 8.3~GHz. As in many AGN, most of
the flux density of M81\* comes from the jet. We estimate that a
compact core must have a flux density of $\simleq 23\%$ of that of the
source, so that the jet to counter-jet ratio is $\simgeq 2.7$.  

In terms of the length of the jet and the estimated accretion rate of
a jet-disk model, M81\* is a scaled-down version of the 
AGN of radio galaxies and quasars, which are several orders of
magnitude more powerful.  The length scale of M81\* is two to three
orders of magnitude shorter and the accretion rate is more than three
orders of magnitude lower. The timescale of ejection for M81\* must be
relatively short, since we show that the energy transport times are
less than a few weeks.  Therefore, in order to keep the source
illuminated, the ejection of new condensations has to happen with a
timescale at least this short if not shorter.  Such a period is at
least an order of magnitude shorter than that observed for more powerful
AGN.  Therefore, observations of M81\* may provide us with a glimpse of the
effectively long-term evolution of an AGN, since we can observe M81\*
over a large number (\ie $>100$) of condensation ejections.  Knowledge
of the long-term behavior, along with the scaling over several orders
of magnitude in size and accretion rate could provide important
additional constraints for models of AGN.

On the other hand, M81\* appears in some respects to be a scaled-{\em
up} version of the nucleus of our own Galaxy, Sgr~A\*.  M81\* has a
radio power four orders of magnitude higher than Sgr~A\*.  Its jet is
much longer, having a length of 400~AU at 43~GHz (extrapolated from
measurements of 700~AU at 22~GHz in Paper~I). At this frequency, Lo
\etal\ (1998) find an intrinsic major axis size of only $\sim4$~AU for
Sgr~A\* (see also Krichbaum \etal\ 1998).
Therefore, the jet in M81\* is at least two orders of magnitude larger
than that of Sgr~A\*.  The accretion rate in M81\* is estimated to be
about the same as that of Sgr~A\* (Narayan \etal\ 1998) or one to two
orders of magnitude larger (see e.g.\ Falcke, Mannheim \& Biermann
1993).

Thus it appears that in terms of power, jet length, and perhaps
accretion rate, M81\* is intermediate between the AGN of radio galaxies
and quasars on the one hand and Sgr~A\* on the other hand, and
therefore investigations of it may help shed some light on the largely
hidden center of our own Galaxy as well.

\section {Conclusions \label{concss}}

\noindent Here we list a summary of our main conclusions:

\begin{trivlist}

\item{1.} A stationary point in the structure of the nucleus of M81
--- identified with the core --- was found south-west of the peak of
the brightness distribution and located with a standard error of about
600~AU.  Over 4 1/2 years, the $2\sigma$ upper bound on its average
velocity in a non-rotating galactic reference frame is 730~\kms\ (or
40\muas~yr$^{-1}$).

\item{2.} A short one-sided jet, variable in orientation and length,
emanates from the core towards the north-east, in projection
approximately along the rotation axis of the galaxy.

\item{3.} The orientation of the jet varies slowly on a timescale of
about 1~year, with an rms of 6\arcdeg\ about the mean of 50\arcdeg\
for the inner part, and 12\arcdeg\ about the mean of 60\arcdeg\ for the
outer part.

\item{4.} The length of the jet varies more quickly.  While some
consistent changes with timescales of $\sim 1$~year are observed, more
rapid fluctuations are also observed.  The mean length of the jet at
8.3~GHz is about 1.0~mas with an rms of 0.2~mas.  Occasionally the jet
appeared somewhat bent counterclockwise.  The changes in both the
length and the position angle of the jet imply that the end of the
visible jet moves with projected velocities, both longitudinal and
transverse, of $\simleq 0.08 c$.  No significant outward motion was
found over the 12 to 18h span of a typical observing run, giving a
$2\sigma$ upper bound on rapid motions of $14 c$.

\item {5.} The flux densities of the dominant component as well as the
total flux density of the source varied erratically, with the time
scale for total flux density variations being less than a few weeks.

\item {6.} Large flux density variations over only a few weeks did
not affect the structure of the source, thus the increase in flux
density must be spread equally throughout the source on a short
timescale. The inferred lower bound on the plasma flow is $0.25 c$.

\item {7.} In one particular case, we can infer a fast plasma flow of
$>0.25 c$, and yet we measure smooth motions of $<0.08 c$ for the end of
the visible jet. This is, to our knowledge, the only case, of
observational evidence for a fast plasma flow and a slow pattern
motion in the same jet at the same time.

\item {8.} The rapid variability of the length and the flux density of
the jet are what is expected from relativistically moving, short-lived
condensations, or condensations in an unsteady flow of plasma, ejected
from the core on timescales less than a few weeks. The slow
variability of the jet orientation points to a possible precession of
the central engine or to slow changes in the orientation of a channel
through which the condensations travel.

\item {9.} The core-jet in M81 may be a scaled-up version of the
largely hidden nucleus in our own Galaxy and could provide a link
between it and the more powerful AGN in radio galaxies and quasars.

\item {10.} From an astrometric point of view, the core in the
nuclear radio source in M81 will become important as a reference
point, with implications for studies of the radio spectrum of the
jet, for the identification of the explosion center of SN1993J
and the ``absolute'' measurement of its expansion velocity, and for
the determination of the proper motion of the gravitational center
of M81 in the extragalactic reference frame and thus the study
of the dynamics of nearby galaxies.

\end{trivlist}

\acknowledgements

ACKNOWLEDGMENTS.  We thank V.I. Altunin, A.J. Beasley, W.H. Cannon,
J.E. Conway, R. Freedman, D.A. Graham, D.L. Jones, A. Rius, G. Umana,
and T. Venturi for help with several aspects of the project.  Research
at York University was partly supported by NSERC. The AIPS
model-fitting routine OMFIT was written by K. Desai.  The NASA/JPL DSN
is operated by JPL/Caltech, under contract with NASA.  We have made
use of NASA's Astrophysics Data System Abstract Service.

\clearpage

\def\phxx{\phantom{xxx}}
\def\baselinestretch{1.1}
\begin{deluxetable}{r@{ }l@{ }l@{\protect\phantom{xxx}} r r r l}
\footnotesize
\tablecaption{8.3~GHz VLBI Observations of M81\* \label{antab}}
\tablewidth{0pt}
\tablehead{
  \colhead{} & \colhead{} & \colhead{} &
  \colhead{} &
  \colhead{Total} &
  \colhead{On-source} &
  \colhead{Recording}\nl
  \multicolumn{3}{c}{Date}   &
  \colhead{Antennas\tablenotemark{a}} &
  \colhead{time\tablenotemark{b}} &
  \colhead{baseline-hrs\tablenotemark{c}} &
  \colhead{mode\tablenotemark{d}}
} 
\startdata
 17&May&93 &
   Eb\phantom{McNtGoRoAqGb}YBrFdHnKp\phantom{La}MkNlOvPtSc &
   \phn 9.6hrs & \phn 10 \phxx & III-B\nl
 19&Sep&93 &
   \phantom{Eb}Mc\phantom{NtGo}RoAq\phantom{Gb}YBrFdHnKpLaMkNlOvPtSc &
   17.6hrs & \phn 55 \phxx & III-B\nl
 6&Nov&93 &
   EbMc\phantom{NtGo}Ro\phantom{AqGb}YBrFdHnKpLaMkNlOvPtSc &
   18.0hrs & \phn 26 \phxx &III-B\nl
 17&Dec&93&
   EbMc\phantom{Nt}Go\phantom{Ro}Aq\phantom{Gb}YBrFdHnKpLaMkNlOvPtSc &
   18.0hrs & \phn 19 \phxx & III-B\nl
 29&Jan&94 &
   Eb\phantom{McNt}GoRoAq\phantom{Gb}YBrFdHnKpLa\phantom{Mk}NlOvPtSc &
   17.6hrs & \phn 23 \phxx & III-B\nl
 15&Mar&94 &
   Eb\phantom{Mc}Nt\phantom{Go}Ro\phantom{Aq}GbYBrFdHnKpLaMkNlOvPtSc &
   18.4hrs & 101 \phxx & III-B\nl 
 22&Apr&94 &
   Eb\phantom{Mc}NtGoRo\phantom{Aq}GbYBrFdHnKpLaMkNlOvPtSc &
   16.9hrs & \phn 59 \phxx & III-B\nl 
 21&Jun&94 &
   Eb\phantom{Mc}NtGoRo\phantom{Aq}GbYBrFdHnKpLaMkNlOvPtSc &
   16.1hrs & 102 \phxx & III-B\nl
 30&Aug&94 &
   Eb\phantom{Mc}Nt\phantom{Go}Ro\phantom{AqGb}YBrFdHnKpLaMkNlOvPtSc &
   14.8hrs & \phn 49 \phxx & III-B\nl
 31&Oct&94 &
   \phantom{Eb}McNt\phantom{Go}RoAq\phantom{Gb}YBrFdHnKpLaMkNlOvPtSc &
   15.1hrs & \phn 56 \phxx & III-B\nl
 23&Dec&94 &
   Eb\phantom{Mc}NtGoRoAqGbYBrFdHnKpLaMkNl\phantom{Ov}PtSc &
   16.1hrs & 106 \phxx & III-B\nl
 12&Feb&95 &
   EbMcNtGoRoAqGbYBrFdHnKpLaMkNlOvPtSc &
   11.8hrs & 190 \phxx & III-B\nl
 11&May&95 &
   \phantom{EbMcNtGoRoAqGb}YBrFdHnKpLaMkNlOvPtSc &
   15.4hrs & \phn 24 \phxx & 128-4-2\nl
 17&Aug&95 &
   \phantom{EbMcNtGoRoAqGb}YBrFdHnKpLaMkNlOvPtSc &
   14.2hrs & \phn 50 \phxx & 128-4-2\nl
 19&Dec&95 &
   \phantom{EbMcNtGo}Ro\phantom{AqGb}YBrFdHnKpLaMkNlOvPtSc &
   15.7hrs & \phn 46 \phxx & 128-4-2\nl
  8&Apr&96 &
   \phantom{EbMcNt}GoRo\phantom{AqGb}Y\phantom{Br}FdHnKpLaMkNlOvPtSc &
   \phn 7.4hrs & \phn 79 \phxx & III-B\nl
  1&Sep&96 &
   Eb\phantom{McNtGoRoAqGb}YBrFdHnKpLaMkNlOvPtSc &
   16.4hrs & \phn 64 \phxx & 128-4-2\nl
 14&Dec&96 &
   Eb\phantom{McNtGoRoAqGb}YBrFdHnKpLaMkNlOvPtSc &
   17.2hrs & \phn 97 \phxx & 256-8-2\nl
  7&Jun&97 &
   Eb\phantom{McNtGoRoAq}GbYBrFdHnKpLaMkNlOvPtSc &
   17.2hrs & \phn 97 \phxx & 256-8-2\nl
 15&Nov&97 &
   EbMcNtGoRo\phantom{Aq}GbYBrFdHnKpLaMkNlOvPtSc &
   13.8hrs & 196 \phxx & 256-8-2\nl
 
\enddata

\tablenotetext{a}{
  Eb= 100m, MPIfR, Effelsberg, Germany;\phn
  Mc=  32m, IdR-CNR, Medicina, Italy;\phn
  Nt=  32m, IdR-CNR, Noto, Italy;\phn 
  Go=  70m, NASA-JPL, Goldstone, CA, USA;\phn
  Ro=  70m, NASA-JPL, Robledo, Spain;\phn
  Aq=  46m, CRESTech/York Univ.\ and Geomatics/NRCan, Algonquin Park, Ontario, Canada;\phn
  Gb=  43m, NRAO, Green Bank, WV, USA;\phn
  Y = equivalent diameter 130m, NRAO, near Socorro, NM, USA;\phn
  Br=  25m, NRAO, Brewster, WA, USA;
  Fd=  25m, NRAO, Fort Davis, TX, USA;
  Hn=  25m, NRAO, Hancock, NH, USA;
  Kp=  25m, NRAO, Kitt Peak, AZ, USA;
  La=  25m, NRAO, Los Alamos, NM, USA;
  Mk=  25m, NRAO, Mauna Kea, HI, USA;
  Nl=  25m, NRAO, North Liberty, IA, USA;
  Ov=  25m, NRAO, Owens Valley, CA, USA;
  Pt=  25m, NRAO, Pie Town, NM, USA;
  Sc=  25m, NRAO, St. Croix, Virgin Islands, USA.
  }
\clearpage
\tablenotetext{b}{Maximum span in hour angle at any one station.}
\tablenotetext{c}{Sum over all the baselines of the number of hours spent on M81\*
after data calibration and editing.}
\tablenotetext{d}{Recording mode: III-B= Mk III mode B double speed; \\
128-4-2= VLBA format, 128 MHz recorded in 4 base-band converters with 2-bit sampling. \\
256-8-2= VLBA format, 256 MHz recorded in 8 base-band converters with 2-bit sampling. \\
  }

\end{deluxetable}

\def\baselinestretch{1.1}
\begin{deluxetable}{r@{\extracolsep{6pt}}l@{}l c 
r@{\extracolsep{3pt}$\>\pm$}l}
\small
\tablecaption{Summary of the VLA Total Flux Density Measurements\label{fluxt}}
\tablewidth{250pt}
\tablehead{
\multicolumn{3}{c}{Date} &
\colhead{Array} &
\multicolumn{2}{c}{Flux Density\tablenotemark{a}~~} \\
  &  &     & Configuration \\
  &  &     &   & \multicolumn{2}{c}{(mJy)~~}
}
\startdata

 17&May&1993 &  B$\rightarrow$C  
                        &114 &\phn 6 \\
 19&Sep&1993 &  DnC     &113 &\phn 6 \\
  6&Nov&1993 &  D       &113 &\phn 6 \\
 17&Dec&1993 &  D       & 85 &\phn 4 \\
 29&Jan&1994 &  D       &119 &\phn 6 \\
 15&Mar&1994 &  A       & 97 &\phn 5 \\
 22&Apr&1994 &  A       &150 &\phn 8 \\
 22&Jun&1994 &  B       &146 &\phn 7 \\
 30&Aug&1994 &  B       &103 &\phn 5 \\
 31&Oct&1994 &  C       &116 &\phn 6 \\
 23&Dec&1994 &  C       & 81 &\phn 4 \\
 12&Feb&1995 &  C       &141 &\phn 7 \\
 11&May&1995 &  D       &134 &\phn 8 \\
 17&Aug&1995 &  A       &176 &\phn 9 \\
 19&Dec&1995 &  B       &144 &\phn 7 \\
  8&Apr&1996 &  C       &165 &\phn 9 \\
  1&Sep&1996 &  D       &161 &\phn 8 \\
 14&Dec&1996 &  A       &164 &\phn 8 \\
  7&Jun&1997 &  AB      &119 &\phn 6 \\
 15&Nov&1997 &  D       &189 &\phn 9 \\
\hline
\multicolumn{4}{c}{Average, rms} & 132 & 30 \\
\enddata
\tablenotetext{a}{The total flux densities of M81\* with estimated standard errors
including both statistical and systematic contributions.}
\end{deluxetable}

\def\baselinestretch{1.1}
\begin{deluxetable}{@{}r@{ }l@{ }l r l@{~~}l@{~~}l@{~}c@{ }l@{~~}l@{~~}l@{~~~}l@{~~}l@{~~}l}
\footnotesize
\tablecaption{Summary of the VLBI model-fitting Results\label{modfitt}}
\tablewidth{0pt}
\tablehead{
& & & &
\multicolumn{3}{c}{Single Component Model} & \protect\phantom{xxx} &
\multicolumn{6}{c}{Two Component Model} \\
\noalign{\vspace{1pt}}
\cline{5-7} \cline{9-14}
\noalign{\vspace{3pt}}
& & & & \multicolumn{3}{c}{Elliptical Gaussian}
    & & \multicolumn{3}{c}{Elliptical Gaussian}
      & \multicolumn{3}{c}{Point Source}\\ 
\noalign{\vspace{1pt}}
\multicolumn{3}{c}{Date} &
\colhead{$S$} &
\multicolumn{1}{l}{~maj}  &\multicolumn{1}{l}{~min}&\multicolumn{1}{l}{p.a.} & &
\multicolumn{1}{l}{~maj}  &\multicolumn{1}{l}{~min}&\multicolumn{1}{l}{p.a.} &
\multicolumn{1}{l}{ratio}&\multicolumn{1}{l}{~~$r$}&\multicolumn{1}{l}{$~~\theta$}
\\
 & & &
\multicolumn{1}{l}{(mJy)} &
\multicolumn{1}{l}{(mas)} &\multicolumn{1}{l}{(mas)} &\multicolumn{1}{l}{~(\arcdeg)} & &
\multicolumn{1}{l}{(mas)} &\multicolumn{1}{l}{(mas)} &\multicolumn{1}{l}{~(\arcdeg)} &
\multicolumn{1}{l}{~(\%) }&\multicolumn{1}{l}{(mas)} &\multicolumn{1}{l}{~(\arcdeg)} 
\\ \hline
\multicolumn{3}{c}{(1)} & \multicolumn{1}{c}{(2)} &
\multicolumn{1}{l}{(3)} & \multicolumn{1}{l}{(4)}&\multicolumn{1}{l}{~(5)} & &
\multicolumn{1}{l}{(6)} & \multicolumn{1}{l}{(7)}&\multicolumn{1}{l}{~(8)} &
\multicolumn{1}{l}{~(9)} & \multicolumn{1}{l}{(10)}&\multicolumn{1}{l}{(11)}
}

\startdata
17&May&93& 114 &$0.44\pm0.03$&$0.05^{+0.04}_{-0.12}$&$59\pm3$& &$0.40\pm0.03$&$0.18^{+0.03}_{-0.09}$&$56\pm3$&$\phn5.3\pm1.0$&$0.56\pm0.05$&$74\pm3$ \\ 
 19&Sep&93 & 113 & 0.46 & 0.16 & 53 & & 0.44 & 0.16 & 53 &\phn2.3 & 0.65 & 80 \\ 
  6&Nov&93 & 113 & 0.51 & 0.21 & 53 & & 0.48 & 0.18 & 49 &\phn5.5 & 0.54 & 81 \\ 
 17&Dec&93 &  85 & 0.44 & 0.21 & 50 & & 0.42 & 0.17 & 46 &\phn4.8 & 0.62 & 82 \\ 
 29&Jan&94 & 119 & 0.46 & 0.16 & 49 & & 0.44 & 0.14 & 46 &\phn3.1 & 0.59 & 75 \\ 
 15&Mar&94 &  97 & 0.62 & 0.17 & 46 & & 0.55 & 0.15 & 40 &\phn8.9 & 0.45 & 65 \\ 
 22&Apr&94 & 150 & 0.61 & 0.14 & 46 & & 0.57 & 0.09 & 40 &\phn9.0 & 0.37 & 70 \\
 22&Jun&94 & 146 & 0.61 & 0.18 & 45 & & 0.48 & 0.16 & 41 &   12.5 & 0.58 & 52 \\ 
 30&Aug&94 & 103 & 0.70 & 0.16 & 47 & & 0.57 & 0.17 & 44 &   11.9 & 0.73 & 52 \\ 
 31&Oct&94 & 116 & 0.41 & 0.13 & 44 & & 0.38 & 0.12 & 45 &\phn8.8 & 0.73 & 48 \\ 
 23&Dec&94 &  81 & 0.48 & 0.21 & 50 & & 0.45 & 0.19 & 48 &\phn8.2 & 0.83 & 48 \\ 
 12&Feb&95 & 141 & 0.44 & 0.19 & 50 & & 0.44 & 0.13 & 50 &\phn6.0 & 0.87 & 45 \\ 
 11&May&95 & 134 & 0.68 & 0.16 & 51 & & 0.57 & 0.14 & 51 &\phn6.3 & 1.07 & 49 \\ 
 17&Aug&95 & 176 & 0.87 & 0.33 & 48 & & 0.74 & 0.13 & 51 &\phn7.9 & 1.00 & 45 \\ 
 19&Dec&95 & 144 & 0.43 & 0.23 & 49 & & 0.40 & 0.17 & 61 &   12.4 & 0.62 & 59 \\ 
  8&Apr&96 & 165 & 0.56 & 0.19 & 57 & & 0.52 & 0.15 & 58 &\phn4.4 & 1.02 & 51 \\ 
  1&Sep&96 & 161 & 0.52 & 0.16 & 46 & & 0.46 & 0.17 & 50 &   10.3 & 0.70 & 55 \\ 
 14&Dec&96 & 164 & 0.48 & 0.27 & 51 & & 0.48 & 0.20 & 58 &   14.4 & 0.86 & 54 \\ 
  7&Jun&97 & 119 & 0.51 & 0.16 & 51 & & 0.47 & 0.16 & 53 &   10.0 & 0.70 & 62 \\ 
 15&Nov&97 & 189 & 0.45 & 0.16 & 55 & & 0.45 & 0.14 & 55 &\phn2.4 & 0.84 & 65 \\ 
\hline\multicolumn{3}{r}{average}
           & 132 & 0.53 & 0.19 & 50 & & 0.49 & 0.16 & 50 &\phn7.7 & 0.72 & 61 \\
\multicolumn{3}{r}{rms} 
           &  30 & 0.11 & 0.05 & \phn4 & & 0.08 & 0.03 & \phn6 &\phn3.4 & 0.18 & 12 \\
\enddata 

\tablenotetext{}{The description of the columns is given below. The estimated
standard errors include statistical and systematic contributions
(see text, \S{\ref{vlbmods}}).\\
(1)  The date (center time of the observing run).\\
(2)  The total flux density of M81\* at 8.3 GHz from simultaneous VLA observations (from 
Table~\ref{fluxt}, repeated here for convenience). \\
(3)  The FWHM of the major axis of the elliptical Gaussian of the one component model.\\
(4)  The FWHM of the minor axis of the elliptical Gaussian of the one component model.\\
(5)  The p.a.\ of the major axis of the elliptical Gaussian of the one component model.\\
(6)  The FWHM major axis of the elliptical Gaussian of the two component model.\\
(7)  The FWHM minor axis of the elliptical Gaussian of the two component model.\\
(8)  The p.a.\ of the major axis of the elliptical Gaussian of the two component model.\\
(9)  The ratio of the flux density of the additional point source to the total flux density.\\
(10) The separation of the additional point source from the elliptical Gaussian.\\ 
(11) The p.a.\ of the position of the additional point source with respect to the 
elliptical Gaussian.
}
\end{deluxetable}

\clearpage

\begin{figure}
\plottwo{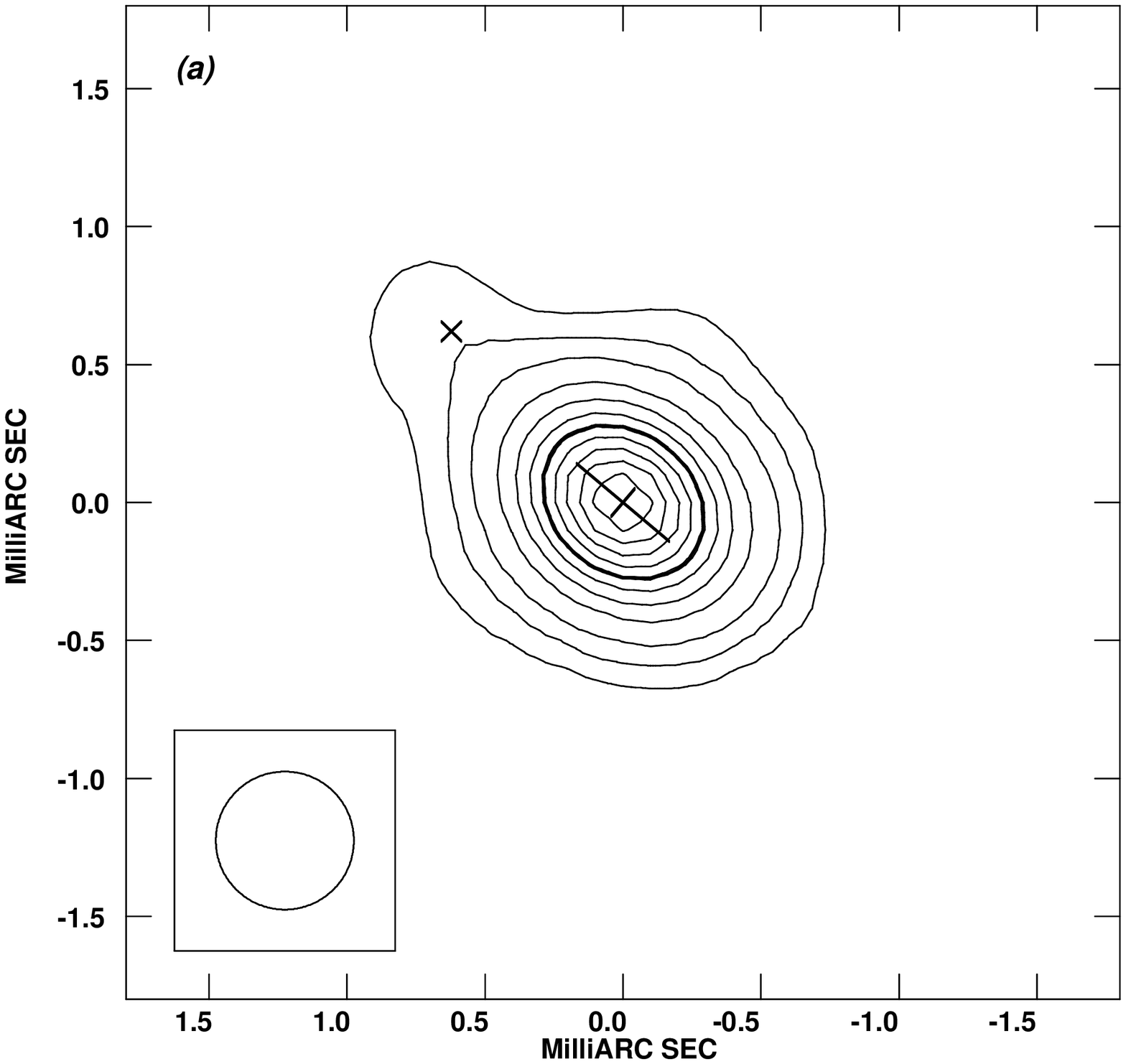}{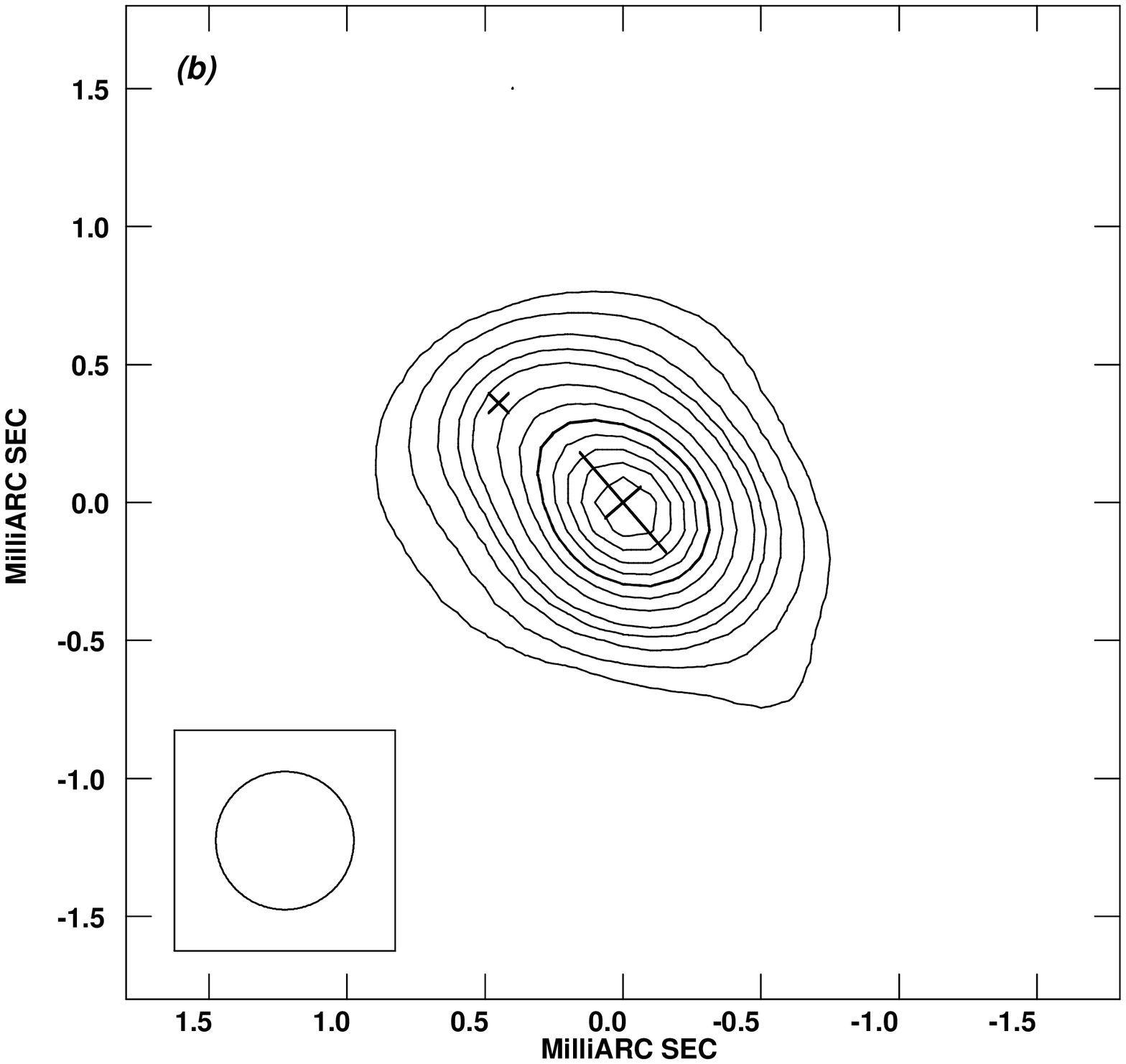}
\end{figure}
\begin{figure}
\vspace{0.1in}
\plottwo{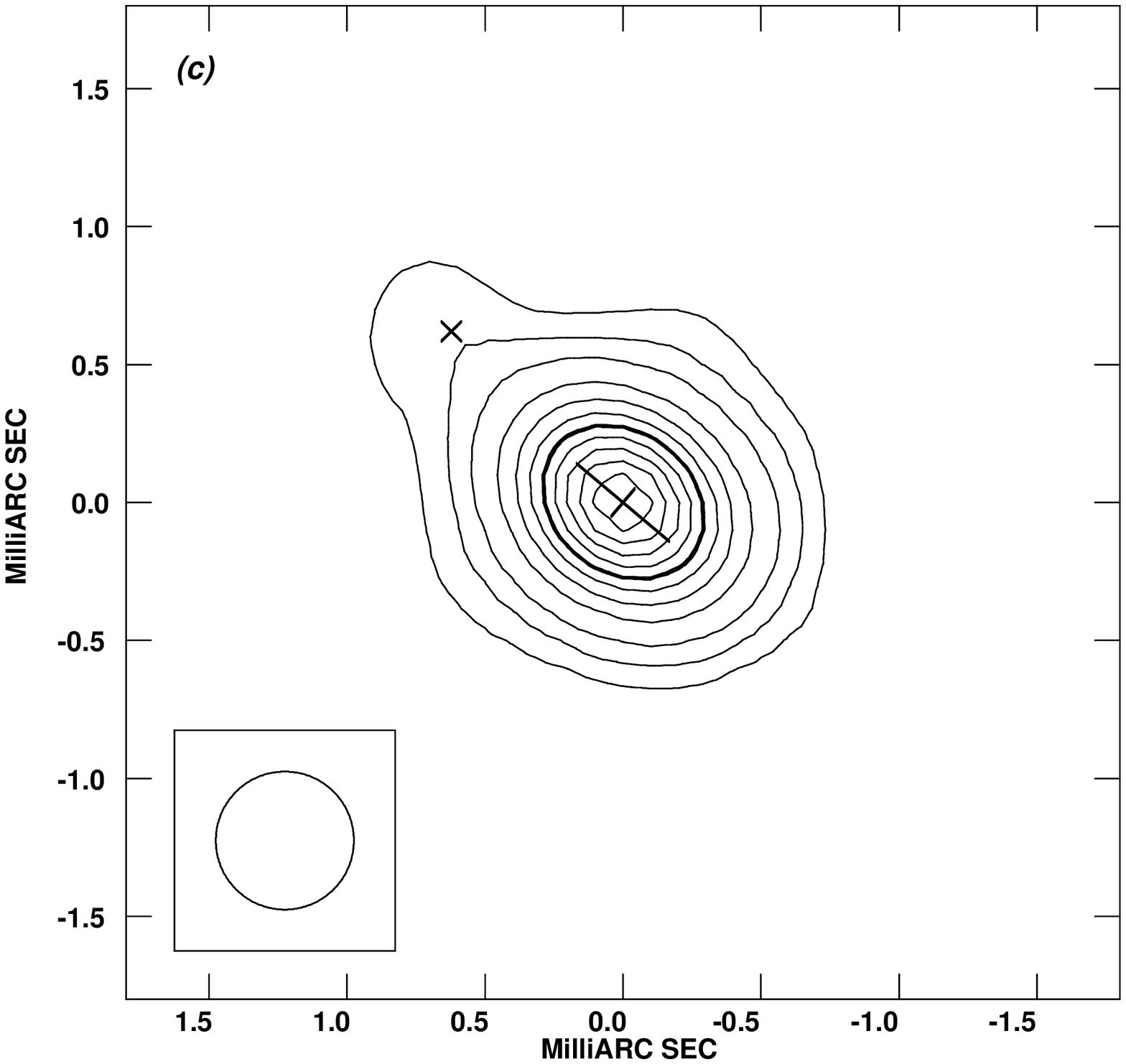}{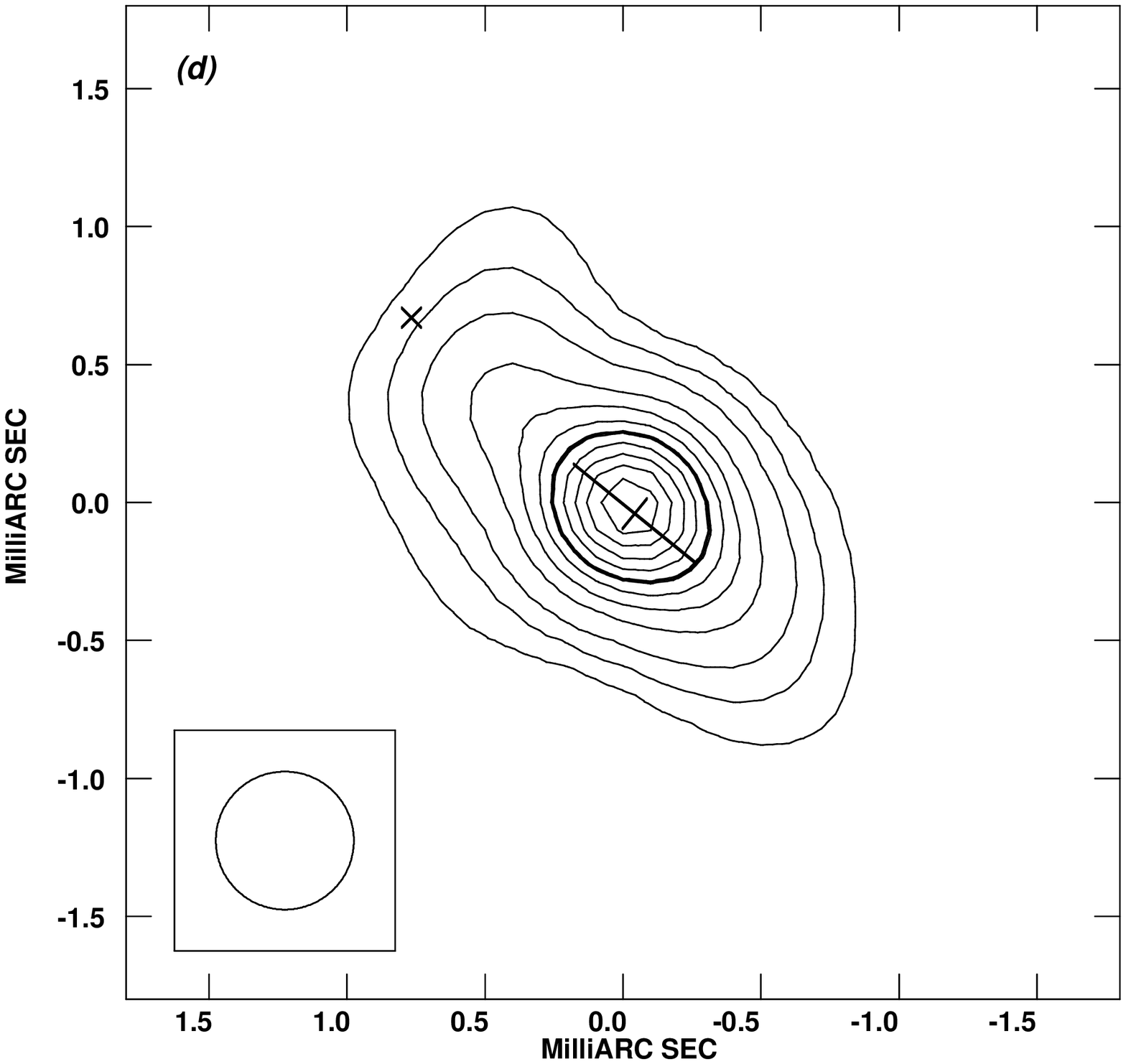}
\end{figure}

\end{document}